\documentclass[aps,prb,twocolumn,groupedaddress]{revtex4}
\usepackage{amssymb,amsmath}
\usepackage{latexsym}
\usepackage{graphicx}
\usepackage{epsfig}
\begin{document}
\title{ Ultrafast Nonlinear Optical Response of
 Strongly Correlated  Systems: Dynamics in the Quantum Hall Effect Regime}
\author{A. T. Karathanos and I. E. Perakis}
\affiliation{Department of Physics, University of Crete,
P. O. Box 2208, 710 03, Heraklion, Crete, Greece}
\author{N. A. Fromer and D. S. Chemla}
\affiliation{Department of Physics,
University of California, Berkeley, CA 94720 and Materials Science
Division, Lawrence Berkeley National Laboratory, Berkeley, CA
94720 }

\date{\today}

\begin{abstract}
We present a  theoretical formulation  of the coherent ultrafast 
nonlinear optical response of a strongly correlated 
system and discuss 
an example where the Coulomb correlations 
dominate. 
We separate out 
the correlated contributions to the third--order 
nonlinear polarization, and  
identify  non--Markovian dephasing effects  coming from  
the non--instantaneous 
interactions and  propagation in time of
the collective excitations of the many--body system. 
We discuss the  signatures, in the time and frequency dependence of the 
four--wave--mixing (FWM) spectrum,  
of the inter--Landau level magnetoplasmon (MP) 
excitations
of the two--dimensional electron gas (2DEG) 
in a perpendicular magnetic field.
 We predict
a resonant enhancement of the lowest Landau level (LL) FWM signal, 
a strong non--Markovian dephasing of the next LL  magnetoexciton (X),
a  symmetric FWM temporal profile, 
and strong oscillations as function of time delay, 
 of quantum kinetic origin. 
We  show that the correlation effects can be controlled 
experimentally 
by tuning the central frequency of the optical 
excitation between the two lowest LLs.
\end{abstract}



\maketitle


\section{Introduction}

The properties
of systems far from equilibrium and, in particular, 
the role of many--body and
collective effects on the femtosecond  and the nanometer scale
present relatively unexplored frontiers of condensed matter
physics.\cite{chemla-01,chemla-99,shah-99,mukamel-book,haugkoch} 
Such problems are particularly challenging in
semiconductors, where the time intervals of interest  are often
shorter than the interaction times and oscillation periods of the
elementary excitations.\cite{chemla-01,chemla-99,weg-00,haugbook}
Examples of well--established pictures
for the interaction processes 
that need be revised in this regime include the semiclassical 
Boltzmann picture
of point-like particles experiencing instantaneous collisions
and the thermal bath
pictures of relaxation and dephasing. \cite{chemla-99,weg-00,haugbook}
Even the notion of weakly interacting ``quasiparticles'', a
cornerstone of condensed matter physics, must be revisited when
describing the ultrafast nonlinear optical response
\cite{chemla-01}.

Wave--mixing experiments are ideally suited for exploring quantum
coherence and collective and correlation effects in semiconductor
nanostructures. \cite{chemla-01,chemla-99,shah-99}
Time--dependent  interactions and correlations 
dominate the FWM signal
during negative time delays, 
where the Pauli blocking effects vanish. \cite{chemla-01,chemla-99}
The treatment of such interactions 
within the time--dependent Hartree--Fock (HF) approximation 
\cite{haugkoch} 
predicts an {\em asymmetric} 
temporal profile 
of the FWM signal.\cite{chemla-01,chemla-99,shah-99} 
The negative
time delay signal generated by  mean field exciton--exciton
interactions decays twice as fast as the  positive time delay
signal. The observation of strong deviations 
from this  asymmetric HF
temporal profile in undoped semiconductors 
was attributed
to exciton--exciton correlations.\cite{chemla-01,chemla-99}

The importance of  many--body effects
in determining 
the time and frequency profile of the 
ultrafast nonlinear  optical spectra 
may be traced microscopically to the 
 coupling, via the interactions, of the
one--particle density matrix that describes the optical
polarization measured in the experiment  to many--particle
correlation functions (e.g. the higher density matrices).
\cite{mukamel-book,chemla-99,axt98}
The latter are factorized within the time--dependent  HF
approximation. \cite{haugkoch} 
The correlation--induced fluctuations, described by the 
deviations from the  factorized form, 
generate 
a new FWM signal, which  can  display a distinct  time and frequency 
dependence as compared 
to the mean field 
signal. Such correlation effects 
are most pronounced  during time scales
shorter than the characteristic times associated with the
interaction processes.
\cite{chemla-99,per-00}

To describe the above non--equilibrium many--body effects,
one must use a
controlled truncation  of the infinite hierarchy of coupled
density matrix or Green function equations. In undoped
semiconductors, this hierarchy truncates if one adopts an
expansion in terms of  the optical fields.
\cite{mukamel-book,axt98,axt-94,axt96,ost-98,per-00} 
This is
the case since 
(a) in the ground state,
the conduction band is
empty and  the valence band is full,
and (b) the Coulomb--induced 
coupling of the conduction and valence bands
via, e.g.,  Auger--like processes  
is  negligible: in the absence of 
optical fields,  the  numbers of conduction band electrons 
and valence band holes are independently conserved.  
In undoped semiconductors, the lowest electronic 
excitations of the ground state electrons 
are the high energy interband $e$--$h$ pairs,
which can adjust almost
instantaneously to the dynamics of the 
photoexcited carriers.
\cite{louie-98} 
The photo-excited $e$-$h$ pairs then behave as
quasi--particles with mutual interactions, while the ground state
can be considered as rigid.
In this case, the many--body nature of the system only affects the
different parameters 
associated with the 
band structure and the dielectric
screening, \cite{sham-66} and the 
only Coulomb correlations that require consideration are
dynamically generated by the optical excitation. \cite{axt98}
The almost unexplored dynamics of 
strongly correlated systems, whose 
ground state electrons 
interact unadiabatically 
with the photo-excited $e$-$h$
pairs, raises very fundamental questions.

A widely used theoretical approach for treating
the above many--body effects in 
undoped semiconductors is
the ``dynamics--controlled truncation scheme'' (DCTS).
\cite{axt98,axt-94,axt96,schaf96}
In this theory, the 
response of the
semiconductor is expanded in terms of the number of created
$e$-$h$ pairs. Importantly,  
the Coulomb  interactions 
that contribute 
to a specified order in the
applied field
only occur between such 
$e$-$h$ pairs. 
This is the case since there is 
the correspondence between the number of 
$e$-$h$ pairs and the sequence of photon absorption and emission, 
and there are no  carriers in the ground state
to interact with the photoexcited carriers. 
The latter condition 
is not met however 
in doped quantum wells, 
where a correlated 2DEG is present in the ground state, 
and the  
DCTS fails there. \cite{axt96} 
A new method that extends the DCTS principles 
to systems with  a strongly correlated ground state is  required. 
In a series of works 
we applied a theory based on a canonical transformation and
time--dependent coherent states  to study the case 
where the interactions between the photoexcited 
$e$--$h$ pairs and the electron Fermi sea (FS) excitations 
dominate the coherent nonlinear optical response.\cite{per-00,per-96-b,prim-00} 

In FS systems, the direct exciton-exciton interactions, which
dominate the nonlinear response in undoped semiconductors,  are
screened, and the nonlinear response is
determined  by the FS excitations. For resonant photoexcitation,
the optical dynamics is dominated by inelastic electron-electron
($e$--$e$) scattering processes.\cite{kim-92,wang-95} At low
temperatures, the dephasing times 
close to the Fermi edge increase by a few picoseconds, in
agreement with Fermi liquid theory.\cite{kim-92} For {\em
below--resonance} excitation, however, the dissipation processes are
suppressed and coherent effects dominate. A novel dynamics of the
Fermi Edge Singularity 
is then observed \cite{per-96-b,bre-95}, due to many-body
correlations of the photoexcited holes with the FS
excitations. \cite{prim-00,per-00,bre-95}

In the absence of long--lived  excitations, a
many--particle system, such as a FS, interacts with 
the photoexcited $e$--$h$ pairs 
almost instantaneously, i.e. during time scales
shorter than the pulse duration. The system then behaves to
first approximation as a thermal bath, and its interactions with
the photoexcited carriers can be treated within the dephasing and
relaxation time approximations. This is not the case however if 
the duration of the interactions
is comparable to or longer
than the measurement times. \cite{prim-00}
In the latter case, the  semiclassical instantaneous collision 
picture breaks down, 
and  quantum mechanical interference effects 
lead to nonexponential decay and 
non--Markovian
memory effects.
\cite{chemla-99,weg-00,haugbook,vu-00}
To  study dephasing in the above  quantum kinetic regime,
one must account for the time
evolution of the {\em coupled} photoexcited carrier--FS
system.\cite{per-00,prim-00}

The change in the energy spectrum caused by 
 a  perpendicular   magnetic field 
restricts the phase space available 
for $e$--$e$ scattering in doped quantum 
wells (QW).\cite{barad-94}
 For strong magnetic fields, 
the Coulomb correlations
are enhanced due to the suppression of the kinetic energy.
\cite{chak-book}
In the  quantum Hall effect (QHE)
regime,\cite{chak-book,QHE2} long lived collective excitations
dominate the 2DEG  spectrum. 
\cite{kallin-84,macd-85-1,fink-97,yusa-01} Recently, the first
experimental studies of the role of  such collective excitations
in the ultrafast nonlinear optical dynamics were
reported. \cite{from-99,from-02,merlin} The presence of low energy
excitations and the strongly correlated ground state 
raise formidable theoretical difficulties for
describing the dephasing dynamics of the 2DEG.

We are  interested in developing a theoretical 
framework  
for describing the ultrafast dephasing and the nonlinear 
optical response of 
strongly correlated systems. 
Examples of  systems of interest 
include  modulation--doped semiconductor QWs, 
where different strongly correlated ground states 
are realized in the QHE regime, 
and the ferromagnetic semiconductors doped with magnetic 
impurities.
In the first part of the paper (sections II--V) 
we  describe the
third--order nonlinear optical response of a many--electron two--band 
system without 
assuming a HF or other specific ground state.
In the second part  (sections VI--VII) 
we study 
the role of the 
inter--LL  MP
collective excitations
in the 
transient FWM spectrum of 
the  cold 2DEG. 
Here we concentrate on  filling factors close to 
$\nu=1$, where 
the spin--$\uparrow$ ground state electrons 
lead to ferromagnetic properties 
(QHE ferromagnet), and the 
excitation spectrum is governed by   
strong Coulomb correlations.\cite{ferro,aif-96,hawr-97}
We consider
photoexcitation with $\sigma_+$ 
circularly polarized light, 
in which case only spin--$\downarrow$ 
electrons are excited
and the MP collective excitations play the most important 
role.\cite{from-02}
Our results explain the most salient qualitative features of the 
transient FWM spectrum observed in  recent experiments.
\cite{from-02}

Our theory   applies to a 
two--band system  described by a Hamiltonian that
independently 
conserves the number of conduction band electrons and valence band 
holes, e.g. the GaAs/AlGaAs QWs.\cite{haugkoch} 
We describe the coupling to the optical 
field  within the dipole approximation, 
and neglect any  stimulated emission. 
We consider  zero 
temperature, 
which is adequate for describing  
correlations that require
thermal energies  smaller than the excitation 
and interaction  energies of the system
in order to be observed. 
The third--order 
polarization calculated here is expected to describe the 
nonlinear optical  signal 
when
the photoexcited carrier density is
smaller than the density of the 
ground state
electrons, in which case  the cold 
2DEG  correlations  prevail.

The outline of the paper is as follows.
In Section \ref{setup} we set up the general problem 
and discuss the nature of the states that contribute 
to the optical spectra.
In 
Section \ref{correl}
we study the time evolution of the 
system, and 
introduce a decomposition of the photoexcited many-body states 
that allows us to classify the different 
interaction contributions. 
In Section \ref{NLPeom} we use the above decomposition 
to derive  the equation of motion for the third--order nonlinear  
 polarization, Eq. (\ref{eom}).
The decompositions introduced in 
Section \ref{correl}
allow us to 
distinguish 
the coherent and excitonic effects from the incoherent 
effects, and separate out the 
factorizable from the correlated 
nonlinear polarization contributions
even in the case of a strongly correlated ground state, 
In Section \ref{dephasing} we 
 discuss an example of  
a basis of strongly correlated states
that can be used to obtain 
equations of motion 
for  the 
correlation functions that describe the 
many--body effects. 
In Section \ref{GPA}
we derive a generalized average 
polarization model, \cite{chemla-99,schaf96,kner-97-99,shah00}  
which we use to  identify the signatures 
of the collective  2DEG excitations
in the time--dependent FWM spectra.
The 
ground state correlations determine 
the interaction parameters 
in the equations of motion. 
We derive in the appendices a number of relations 
among such interaction parameters that are  
imposed by the electron--hole symmetry 
of the ideal 2DEG. 
In Section \ref{exper} 
we 
present numerical results that describe the correlation--induced 
ultrafast dynamics predicted by the above model. 
We identify a number of  interesting 
features 
in the time--dependent FWM spectrum, which arise 
from the propagation in time of the inter--LL 
MPs and their non--instantaneous interactions  with the photoexcited excitons. 
We end with the conclusions.

\section{Problem setup }

\label{setup}

We are interested in developing a comprehensive approach to the
problem of the nonlinear optical response 
in the case of 
photoexcitation from the valence to the conduction band. 
Within the dipole approximation, the coupling 
to the optical field 
can be
described by the Hamiltonian \cite{haugkoch} ($\hbar=1$)
\begin{equation} \label{H-tot}
H_{{\rm tot}}(t) = H -\mu {\cal E}(t) {\hat X}^{\dag}- \mu {\cal
E}^*(t) {\hat X}.
\end{equation}
In the above equation, 
$H$ is the ``bare''  many--body Hamiltonian, 
which describes the bandstructure effects and the interactions,
 ${\cal E}(t)$ is the applied optical field, $\hat{X}$ 
is the optical transition operator, and $\mu$ 
is the interband transition matrix element.
In the  case of a 
semiconductor QW
containing a 2DEG in a  magnetic field, 
the Hamiltonian $H$ has the form
\cite{haugkoch}
\begin{eqnarray} 
H =  \sum_{i, k} [E_g + \Omega_c^c (i + 1/2)] \, 
 {\hat
e}^{\dag}_{k,i} {\hat e}_{ k,i}  \nonumber \\ +
\, \sum_{i, k}  \Omega_c^v(i + 1/2) \,
{\hat h}_{- k,i}^{\dag} {\hat
h}_{-k,i} \, 
+ V_{ee} +V_{hh}+V_{eh}, \label{H-el}
\end{eqnarray}
where $E_{g}$ is the bandgap, and $V_{ee}, V_{eh}$, and $V_{hh}$ are,
respectively, the {\em e--e}, {\em e--h}, and {\em h--h}
interactions (see Appendix \ref{symm}).
 The magnetic field splits
the conduction and valence bands into 
discrete electron ($e$) and hole
($h$) LLs, $e$--LLi and $h$--LLi, 
where $i$ includes both the LL index and the spin. 
${\hat e}^{\dag}_{k,i}$ is the creation operator of the LL$i$ conduction band
electron,
with cyclotron energy $\Omega_c^c$, and 
${\hat h}^{\dag}_{ k ,i }$ is the creation operator of the
LL$i$ valence band hole, with cyclotron energy $\Omega_c^v$
(see Appendix \ref{symm}). \cite{macd-85-1} 
The optical transition operator
${\hat X}^{\dag}$
is expanded in terms of 
interband $e$--$h$ pair creation 
operators ${\hat X}_i^{\dag}$ that we refer to 
as  the exciton (X) operators from now on: 
\begin{equation}
\label{X}
{\hat X}^{\dag} =
\sum_i \sqrt{N_i}\, {\hat X}_i^{\dag}.
\end{equation}
In the case of the 2DEG in a magnetic field 
it is convenient to 
introduce the  LL$i$
magnetoexciton states 
$|X_i\rangle=\hat{X}^\dag_i|0 \rangle$, 
where $|0\rangle$ is the ground eigenstate of the 
many--body Hamiltonian $H$, 
with full valence band and the  2DEG at rest.
The eigenvalue equation $H | 0 \rangle = 0$ defines the ground 
state energy as the 
reference point.
In the ideal system, 
\begin{equation}
\label{Xi-def} 
{\hat X}_i ^\dag = \frac{1}{\sqrt{N_i}} \sum_{ k} {\hat
e}^{\dag}_{ k,i} {\hat h}^{\dag}_{ -k,i},
\end{equation} 
where  $N_i=N (1-  \nu_i)$, with 
$N =L^2/2 \pi l^2$ being the LL
degeneracy, $l$ the magnetic length, $L$ the system size, and
\begin{equation} 
\nu_i= \frac{1}{N} \sum_{k} \langle 0| 
{\hat e}_{k,i}^{\dag} {\hat e}_{ k,i} | 0 \rangle 
\label{fill} 
\end{equation} 
gives the filling of LL$i$  in the absence of optical excitation.
Note that
 the exciton states $| X_i \rangle$
  are 
strongly  correlated: they are 
created by the operator $\hat{X}^\dag_i$ 
acting on the ground eigenstate of the 
 many--body Hamiltonian $H$, which describes the 
correlated electron gas at rest. 
From 
Eq. (\ref{Xi-def}) we obtain 
the
commutation relation
\begin{equation}
\label{commut} [{\hat X}_i, {\hat X}_j^{\dag}] = \delta_{ij}
\left( 1 - \frac{\Delta \hat{N}_i}{N_i} \right),
\end{equation}
where  the operator 
\begin{equation}
\label{DN} \Delta \hat{N}_i = 
\sum_{ k} \left( {\hat
h}_{- k,i}^{\dag}  {\hat h}_{- k,i} + 
{\hat e}_{k,i}^{\dag} {\hat e}_{ k,i} \right) - N \nu_i,
\end{equation}
with $\langle 0 | \Delta \hat{N}_i | 0 \rangle =0$, 
describes the 
number of  photoexcited carriers in LLi.

The optical spectra are
determined by the polarization of the photo-excited system,
\begin{equation} \label{Ptot}
P(t) = \mu \langle \psi | {\hat X} | \psi \rangle = \mu \sum_i
\sqrt{N_i} P_i(t),
\end{equation}
where $P_i$ are the average values of the exciton operators 
\begin{equation} \label{Pi}
P_i(t) = \langle \psi | {\hat X}_i | \psi \rangle.
\end{equation}
The state $| \psi(t) \rangle$ 
evolves from the 
state of the system prior to the optical excitation
according to the
Schr\"{o}dinger equation for the Hamiltonian $H_{tot}(t)$. 
For zero temperature,   this initial 
state is the lowest many--body eigenstate $| 0 \rangle$
and  describes all correlations in the absence 
of  optical fields.

As in the theoretical approaches of Refs.\onlinecite{axt96,ost-98},
there is a one to one correspondence
between the photon absorption/emission  and the $e$-$h$
pair creation/destruction. Since an electron gas 
may be  present
in the ground state, 
we classify  the photoexcited states 
in terms of the  number of valence band
holes, i.e. the number of  missing valence band electrons 
as compared to the ground state $|0\rangle$.
We thus decompose the optically--excited
state $|\psi\rangle$ as
\begin{equation} \label{psiexpansion}
| \psi \rangle = | \psi_0 \rangle + | \psi_1 \rangle + | \psi_2
\rangle,
\end{equation}
where $| \psi_n \rangle$ is the 
{\em collective} $n$--$h$ 
 photoexcited state. 
The above holes interact strongly with the 2DEG.\cite{hawr-97}
Note that the states with $n \ge 3$ do not 
contribute to the third--order nonlinear polarization.
\cite{mukamel-book}

\begin{figure}[h]
\begin{center}
\includegraphics*[width = 9cm,height=4cm]{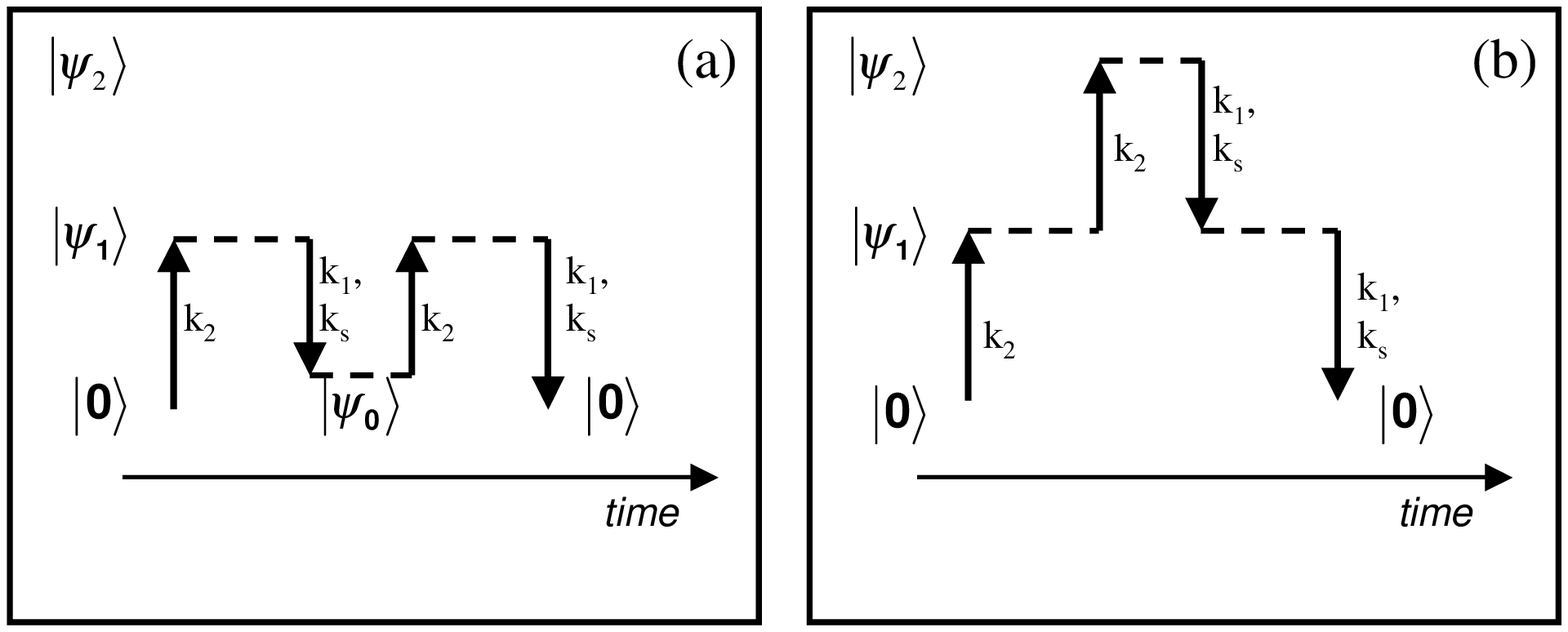}
\caption{
Photoexcitation of the 
intermediate (a) 0--$h$, $| \psi_0 \rangle$, 
and (b) 2--$h$,  $| \psi_2 \rangle$,  
states 
via the nonlinear optical processes
that contribute to the FWM spectrum. 
To third order in the optical fields, 
the coherent emission 
of a $k_s$ photon 
in the FWM direction 
$k_s= 2 k_2 - k_1$ 
is determined by the excitation of two 
$e$--$h$ pairs by the optical field 
$k_2$, and the deexcitation 
of one $e$--$h$ pair by the optical field $k_1$.
Although in a coherent FWM experiment 
we must begin and end the nonlinear excitation 
process with the system in its ground state, 
the intermediate 0--$h$ state does not need to be the 
ground state $|0\rangle$, but can contain 
electron gas excitations. The above optical transitions are assisted 
by Coulomb interactions, which lead to the correlations 
discussed in Section \ref{correl}.  
}
\label{FWM-process}
\end{center}
\end{figure}

Substituting Eq.(\ref{psiexpansion})  into the
Schr\"{o}dinger equation for the Hamiltonian $H_{tot}(t)$ we
obtain up to third--order in the optical field that
\begin{eqnarray}
i \partial_t | \psi_0 \rangle - H | \psi_0 \rangle = - \mu {\cal
E}^* {\hat X} | \psi_1 \rangle,
\label{psi0}\\
i \partial_t | \psi_1 \rangle - H | \psi_1 \rangle = - \mu {\cal
E} {\hat X}^\dag | \psi_0 \rangle - \mu {\cal E}^* {\hat X} |
\psi_2 \rangle,
\label{psi1}\\
i \partial_t | \psi_2 \rangle - H | \psi_2 \rangle = - \mu {\cal
 E} {\hat X}^\dag | \psi_1 \rangle
\label{psi2}
\end{eqnarray}
with initial condition $| \psi_n(-\infty) \rangle = \delta_{n,0} |
0 \rangle$, where 
the Hamiltonian $H$ includes the degrees of freedom 
that lead to the dephasing. 
 The physics of the above equations is clearly
displayed: $|\psi_0 \rangle$ is coupled to $|\psi_1 \rangle$ by
the destruction of one $e$-$h$ pair, $|\psi_1 \rangle$ is coupled
to $|\psi_2 \rangle$ by the destruction of one $e$-$h$ pair and to
$|\psi_0 \rangle$ by the creation of one $e$-$h$ pair, and
$|\psi_2 \rangle$ is coupled to $|\psi_1 \rangle$ by the creation
of one $e$-$h$ pair. 
Fig. \ref{FWM-process} shows the 
optical transitions that determine the 
FWM signal up to third order in the optical field.
It is worth noting that, by retaining in the
expansion Eq. (\ref {psiexpansion}) states with higher $h$
numbers, one can extend Eqs. (\ref{psi0}-\ref{psi2}) to treat
higher order nonlinear processes.

Even if we restrict ourselves to the electronic degrees of freedom, 
the Hilbert space of states that determine 
the ultrafast nonlinear response of a doped QW 
is  complicated. Strictly
speaking it contains all the states that can be generated through
the coupling of $e$-$h$ pairs photoexcited in any of the QW
subbands with all the excitations of the 2DEG: plasmons, magnons,
incoherent pairs, etc. For the purpose of 
developing an intuitive picture of the
important 
physical 
processes it is useful
to first discuss qualitatively the ensemble of states that are most relevant
to the problem at hand.

For the experimental conditions  considered in the second 
part of the paper, the most
important 2DEG excitations are the collective inter--LL 
MP 
modes, which arise from
the coherent promotion of a LL0 electron to a higher
LL. \cite{kallin-84,macd-85-1} Such MP eigenstates 
are well approximated by the form \cite{chak-book} 
\begin{equation}
\label{MP} |M_{{\bf q}} \rangle = \sum_{ k  j j'} \rho_{jj'}({\bf
q}) {\hat e}_{ k + q_y,j}^\dagger {\hat e}_{ k,j'} |0
\rangle \,,
\end{equation}
where $| 0 \rangle$ is the strongly correlated ground state 
and the amplitudes
$\rho_{jj'}({\bf q})$ are related to the LL$j'$ $\to$ LL$j$
contribution to the density operator. 
Note that, similar to the exciton states $|X_i\rangle$, 
the above MP states are strongly correlated. 
For the
magnetic fields of interest, in the  ground
state $| 0 \rangle$, 
only e--LL0 is partially filled with the 2DEG at rest, 
while all  the $h$-LL states are empty (full valence band). 
Since we focus on photoexcitation of  the LL0 and LL1 optical
transitions, the main contribution to the optical spectra comes
from the resonant LL0 $\to$ LL1 MPs,
 whose energy is close to the LL0 $\to$ LL1 energy.
\cite{kallin-84,macd-85-1} 

It is useful to make the junction with two domains well studied in
the recent literature: photoexcited undoped QW, and 
 2DEG in the QHE
regime. One can distinguish 
between the excitations of two subsystems:
(i) the QW interband excitations 
(with the 2DEG at rest), which consist of
1$e$-$h$, 2$e$-$h$, $\cdots$ pairs  created in the
different QW LLs, and (ii) the 2DEG excitations (with unexcited
QW and full valence band), 
i.e. the  1-MP, 2-MP, $\cdots$  states, etc.
The ensemble of
states that determine the third--order nonlinear optical spectra
can then be thought as consisting of 
$\ell$$e$-$h$ pairs, $l \le 2$, 
 and n 2DEG excitations. 
For photoexcitation of the LL0 and LL1 
exciton transitions, 
the  inter--LL MP provides a resonant 
coupling of the two LLs, since its energy  
is comparable to the LL0 $\rightarrow$ LL1 excitation energy. 
In contrast,  
the  LLi exciton states with 
$i \ge 3$,
the  states with $n \ge 2$ 
MPs,  and the continuum of 
incoherent 2DEG pair 
excitations analogous to the ones 
in an ordinary Fermi liquid \cite{chak-book} 
primarily contribute to the optical spectra 
via non--resonant 
processes. 

One can draw an analogy between the
X--MP effects of interest here and  the X--phonon interaction
effects studied in undoped semiconductors.
\cite{weg-00,haugbook,axt98,axt96,vu-00,schilp-94} However, there
are some important differences. 
In the undoped system, 
the electronic operators commute
with the collective 
excitation (phonon) operators, 
and the ground state correlations can be 
neglected. 
One can then expand the state $|\psi\rangle$ in terms of a basis
consisting of products of phonon wavefunctions times
$e$-$h$ pair two--particle wavefunctions.
 In contrast, a MP is an electronic 
excitation (see Eq.\ (\ref{MP})), and its creation operator may not
commute with other electronic operators.  Pauli exchange
effects must then be considered, while, unlike for
phonons, MPs do not strictly obey Bose statistics.
Importantly, 
one must 
treat the strong correlations of the 
ground state 
electrons.
Issues such as the above
complicate the use of a simple basis to calculate the nonlinear
optical response of the 2DEG. 
In Section \ref{dephasing}
we discuss an example 
of a strongly correlated basis set 
that can used to address the 
above issues.
An important  advantage of this
particular  basis is that it facilitates  the  development of  
a simple model 
that describes  the most salient 
dynamical features of the ultrafast 
nonlinear optical spectra.

\section{Time--dependent interaction effects}

 \label{correl}

In this section we consider the time evolution of the
coupled photoexcited carrier--2DEG system that
 leads to the dephasing of the $e$--$h$ polarization.
We are  mainly interested in  dephasing due to 
electronic degrees of freedom,
and thus the distinction 
between the photoexcited carriers and the 
``bath'' excitations   is less 
clear as compared e.g. to the case of a phonon bath. 
We address this issue by separating out 
the excitonic contribution directly excited by the optical field 
(2DEG at rest) from 
the contribution of the excited 2DEG configurations
(denoted by 2DEG$^*$ from now on) that lead to the dephasing. 
For this we decompose the 1--$h$ photoexcited state 
as follows: 
\begin{equation}
\label{1hole} | \psi_1 \rangle = \sum_i P_{i}^{L} | X_i \rangle +
| \bar{\psi}_1 \rangle,
\end{equation}
where  $|\bar{\psi}_1\rangle$ is the 
\{1-$h$/2DEG*\} contribution defined by the condition 
$\langle X_i|\bar{\psi}_1\rangle=0$, 
and the exciton amplitude 
\begin{equation}
\label{pol} P_i^{L} = \langle X_i | \psi_1 \rangle = \langle 0 |
{\hat X}_i | \psi \rangle
\end{equation}
reduces to the 
linear  polarization to first order in the optical field.  
To describe the two--photon nonlinear optical processes in 
Fig. (\ref{FWM-process}), 
we must consider, in addition to the 
X--2DEG interactions, the 
X--X and X--$|\bar{\psi}_1\rangle$
interactions during the optical transitions. 
For this we 
first separate out the total interaction  contribution to the
2-$h$ and 0-$h$ intermediate 
states, 
 $|\psi_2\rangle$ and
$|\psi_0\rangle$ respectively,
and then identify the particular 
contributions 
due to the  interactions
among the above  1--$h$ excitations that lead to 
the correlation 
effects: 
\begin{eqnarray}
 | \psi_2\rangle =  \frac{1}{2} \sum_{ii'} P_{i}^L
P_{i'}^L | X_{i} X_{i'} \rangle + | \psi_2^{int}\rangle, 
\nonumber 
\\
|\psi_2^{int}\rangle = \sum_{i} P_{i}^L {\hat X}^{\dag}_{i} |
\bar{\psi}_1 \rangle + |\bar{ \psi}_2\rangle, 
\label{2hole}
\end{eqnarray}
where the state 
$| X_i X_{i'} \rangle = {\hat
X}^\dag_i {\hat X}^\dag_{i'} | 0 \rangle$ 
describes two non--interacting
Xs,
and 
\begin{eqnarray}
| \psi_0 \rangle = \langle 0 | \psi \rangle \, | 0
\rangle + | \psi_0^{int} \rangle \ , \ \nonumber \\
 | \psi_0^{int} \rangle = -
\sum_{i} P_{i}^{L*} {\hat X}_{i} | \bar{\psi}_1 \rangle + |
\bar{\psi}_0 \rangle,
\label{0hole} 
\end{eqnarray}
where 
we have  separated out the  ground state contribution 
from the 2DEG$^*$ contributions  
by requiring that 
$ \langle 0|\psi^{int}_0 \rangle = \langle 0|\bar{\psi}_0
\rangle  = 0$.

 The above decompositions are analogous
to the 
cumulants introduced within the DCTS 
for the case of undoped semiconductors. 
\cite{axt98,axt96}
Such cumulants were 
obtained by subtracting the factorized contributions from the
many--body correlation functions.
 Note however that 
the  method 
presented below also
holds in the case where strongly correlated 
carriers are present in the ground state,
as in the 2DEG case, where the assumptions of the DCTS break
down.
The decompositions of the states $|\psi_0^{int}\rangle$,
$|\psi_2^{int}\rangle$, and $|\psi_1\rangle$ also 
allow us to separate out, 
 in the equations of motion Eqs.\ (\ref{psi0}),
(\ref{psi1}), and (\ref{psi2}), the source terms proportional to
the optical field from the source terms proportional to the
polarizations $P_i^L$, which lead to different time dependencies.
The photo-excited states 
$|\bar{\psi}_0 \rangle$, $| 
\bar{\psi}_1\rangle$, and
$|\bar{\psi}_2\rangle$
describe correlated contributions, 
whose physical origin will be discussed below.

We now derive the equations of motion 
of the above photoexcited states, 
which we will use in the next section 
to derive the nonlinear polarization equation of motion and 
separate 
out the factorizable contributions.
It is easiest to start with the 1--$h$ time--evolved state.
Eq.\ (\ref{1hole}) 
splits this  state  into
 excitonic (2DEG at rest)   and  
\{1-$h$/2DEG*\} parts, $P_i^L(t)$ and
$|\bar{\psi}_1\rangle$ respectively. 
The state $| \bar{\psi}_1 \rangle$
originates
from the  $X$--2DEG scattering during the time evolution of the
photo-excited X.
To describe such interactions, we consider
the action of the Hamiltonian $H$ on the
exciton states $|X_i \rangle$. By subtracting  all  the
exciton contributions, the state $H | X_i \rangle$ can be
expressed in the  form
\begin{equation}
H |X_i \rangle = \Omega_i |X_i \rangle -  \sum_{i' \neq i} V_{i'i}
|X_{i'} \rangle + |Y_i \rangle \,, \label{HonX}
\end{equation}
where
\begin{equation}
\label{Xen} \Omega_i = \langle X_{i}| H | X_i \rangle
\end{equation}
is the $X_i$ energy,
\begin{equation}
\label{V} V_{i'i} = -\langle  X_{i'}| H | X_{i} \rangle =
V_{ii'}^{*}
\end{equation}
describes the Coulomb--induced coupling of the different 
Xs, and $|Y_i \rangle = {\hat Y}_i ^\dag |0
\rangle $, where the operator
\begin{equation}
\label{Yop} {\hat Y}_i = [{\hat X}_i,H] - \Omega_i {\hat X}_i +
\sum_{i' \neq i} V_{ii'} {\hat X}_{i'},
\end{equation}
describes the interactions between
$X_i$ and the rest of the carriers present in the
system.

\begin{figure}
\begin{center}
\includegraphics*[width = 8.5cm,height=6cm]{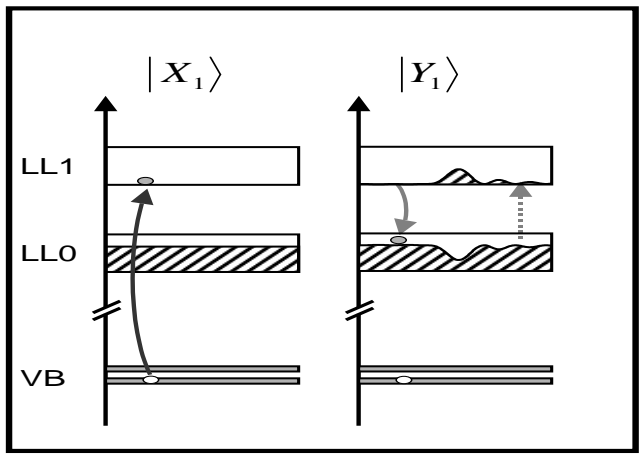}
\caption{
Photoexcitation of the 
exciton state, $| X_1\rangle$, 
and then of the X+MP state $|Y_1 \rangle$ 
via resonant X--2DEG scattering.
}
\label{Y-cartoon}
\end{center}
\end{figure}

As one can see  by using the above equations, 
the state $|Y_i \rangle$
is orthogonal to all  exciton states $| X_j
\rangle$, $\langle Y_i | X_j \rangle =0$,
and is therefore  the \{1-$h$/2DEG*\}  state
into which $X_i$ can scatter 
by interacting with the 2DEG. 
For the experimental conditions of particular interest here, 
the most important contribution 
to $|Y_i\rangle$, Eq. (\ref{Y-symm}), 
comes from  X+MP states. 
To see this, 
let us consider the 
possible final scattering states of the 
LL1 exciton $X_1$.
Its LL1  electron can scatter to 
LL0 by emitting a 
LL0 $\to$ LL1 MP,
a process shown in Fig. \ref{Y-cartoon}. 
Since the MP 
energy is  close to the 
$e$--LL0 $\to$ $e$--LL1 energy spacing,  
the above scattering  process is almost resonant. 
It therefore 
 provides an efficient 
decay channel of the LL1 exciton 
to a 
\{1-MP + 1-LL0-e + 1-LL1-h\} four-particle excitation
of the ground state 
$| 0 \rangle$.
All other allowed scattering 
processes are nonresonant. 
The $X_1$  hole can scatter to 
LL0 by emitting a MP, which leads to
a \{1-MP + 1-LL1-e + 1-LL0-h\} four--particle excitation.
The latter state however has energy that is
significantly higher, by an amount 
of the order of $\sim \Omega_c^c+ 
\Omega_c^v$,  from that of 
the initial $X_1$ state.
Note that, as shown in Appendix \ref{symm}, 
in the electron--hole symmetric limit 
the X electron or hole must change LL 
during the scattering process. 
In the case of $X_0$, 
the LL0 electron can scatter to LL1
by emitting a MP, 
so that 
$X_0 \rightarrow$ \{1-MP + 1-LL1-e + 1-LL0-h\},
or the LL1 hole can scatter to LL0, 
in which case $X_0 \rightarrow$ 
\{1-MP + 1-LL0-e + 1-LL1-h\}.
$|Y_0\rangle$ is thus a linear combination 
of the same final states as $|Y_1\rangle$.  
However, in this case the energy of all final states 
is   significantly higher than that of the  
initial state $|X_0\rangle$. Therefore, 
the decay of the LL0 exciton 
is suppressed as compared to that of 
the LL1 (or higher) exciton.
Note that the distinction between resonant and nonresonant 
interaction  processes
is most pronounced 
when the 
inter--LL 
excitation energy, of the order of the cyclotron 
energy, exceeds the 
characteristic 2DEG Coulomb correlation energy, 
$\sim e^2/l$.
Expansions in terms of the ratio of the above 
two energies are known to capture  most of the 
2DEG correlation effects. \cite{chak-book,kallin-84,hawr-97}

We  now describe the time evolution, to 
first order in the optical field, 
 of the 1--$h$
photo-excited state $| \psi_1 \rangle$. 
The equation of motion 
for the linear polarization 
$P_i^L$ can be derived by 
projecting the exciton state $\langle X_i |$ 
to the 
truncated Eq.\ (\ref{psi1}) 
and applying Eq.\ (\ref{HonX}):
\begin{equation}
\label{linpol} i \partial_t P_{i}^{L}
 = \Omega_i  P_{i}^{L}
- \sum_{i' \ne i} V_{ii'} P_{i'}^{L} + \bar{P}_{i}^{L} - \mu {\cal
E}(t) N_i^{1/2}.
\end{equation}
The correlation function
\begin{equation}
\label{barpol} \bar{P}_i^{L} = \langle Y_i |\psi_1 \rangle =
\langle Y_i |\bar{\psi_1} \rangle,
\end{equation}
 discussed in Section \ref{dephasing},
describes the dephasing of $P_i^L$ and screening effects. 

Substituting the decomposition Eq.\ (\ref{1hole}) into the
Schr\"odinger equation Eq.(\ref{psi1}), and using Eqs.
(\ref{linpol}) and  (\ref{HonX}), we obtain the equation of
motion of the \{1-$h$/2DEG*\} contribution $|\bar{\psi_1}
\rangle$:
\begin{equation} \label{bar1hole}
i \partial_t |\bar{\psi_1} \rangle - H |\bar{\psi_1} \rangle =
\sum_{i} \left[ P_i^L |Y_i \rangle
- \bar{P}_i^L | X_i \rangle \right].
\end{equation}
The operator $P_i^L \hat{Y}^\dag_i - \bar{P}_i^L \hat{ X}^\dag_i$
also appears in the equation of motion of the 
2h state. Its first term
 describes the scattering of $X_i$
 with the 2DEG, while 
its second term compensates for the dephasing of
$P_i^L$ and ensures the orthogonality $\langle X_i |
\bar{\psi}_1 \rangle =0$.

We can perform a similar analysis of the time--evolved 2--$h$
state by separating 
out 
in Eq. (\ref{2hole}) 
the contribution of the non--interacting
two--exciton states $| X_i X_{i'} \rangle$.
\cite{shah00,ost-98}
This contribution  describes the
time evolution of the two Xs photo-excited by the
optical field in the absence of interactions. However, the two
Xs interact with each other as well as with the 2DEG, as
described by the equation
\begin{eqnarray}
H | X_i X_{i'} \rangle = (\Omega_{i} + \Omega_{i'}) |  X_{i}
X_{i'} \rangle - \sum_{j \ne i'} V_{ji'} | X_{i} X_{j} \rangle 
\nonumber  \\ 
- \sum_{j \ne i} V_{ji} | X_{j}  X_{i'}  \rangle
+ | X_i  Y_{i'} \rangle +  | X_{i'} Y_i \rangle + | B_{ii'}
\rangle, \label{HonXX}
\end{eqnarray}
obtained by using Eq.\ (\ref{Yop}) to calculate the state
$[H,{\hat X}^\dag_{i}{\hat X}^\dag_{i'}]|0 \rangle$. The first
term in Eq.\ (\ref{HonXX}) is the energy of the two
non--interacting $X$s, while the next two terms come from the
Coulomb--induced LL coupling.
Similar to $|X_i X_{i'} \rangle$, the state $| X_i
Y_{i'} \rangle = {\hat X}_i^\dag {\hat Y}^\dag_{i'} | 0 \rangle$
describes a noninteracting pair of 
$X_i$ and  $Y_{i'}$  excitations.
Finally, as for the undoped case, the last term in
Eq.\ (\ref{HonXX}),
\begin{equation}
\label{B} |  B_{ii'} \rangle = [{\hat Y}^\dag_i , {\hat
X}^{\dag}_{i'}] | 0 \rangle = [[H, {\hat X}^{\dag}_{i}], {\hat
X}^{\dag}_{i'}] | 0 \rangle,  
\end{equation}
comes from the X--X interactions.
\cite{ost-98,shah00}
 Eq. (\ref{B-symm}) demonstrates 
that the state 
$|B_{ii'}\rangle$ 
is a linear combination of two {\em
e--h} pairs with different center of mass momenta, 
but with the 2DEG
in its ground state, and thus describes
biexciton bound, $X_2$, and scattering, $XX$, states similar to the 
undoped system.\cite{ost-98,shah00} 

 The $X$-$X$ and $X$-2DEG interactions
contribute to the time evolution of the photo-excited 2--$h$ state
in Eq.\ (\ref{2hole})
through $|\psi_2^{int} \rangle$. We further decompose 
the latter state into
(a) the contribution of the non--interacting 
pair of  $X_j$--$|\bar{\psi}_1\rangle$
1$h$ excitations, and (b) the contribution
$|\bar{\psi}_2 \rangle$ due to the  interactions between all the
different pairs of 1--$h$ excitations, i.e. the 
$X$--$X$ interactions (as  in the undoped system) 
and the  
$X$
interactions with the \{1-$h$/2DEG*\} states
(such as the four-particle Y excitations discussed
above).

To obtain the equation of motion of
the correlated 2--$h$ contribution  $|\bar{\psi}_2 \rangle$,
we note that the 
time-evolved  state $| \psi_2 \rangle $
contributes to the
third--order nonlinear 
response  at second order in the applied field. By
taking the time derivative of Eq.\ (\ref{2hole}) and using Eqs.\
(\ref{psi2}),  (\ref{Yop}), (\ref{linpol}), 
(\ref{bar1hole}), and (\ref{HonXX}), we obtain that 
\begin{eqnarray}
 i \partial_t | \bar{\psi}_2 \rangle - H |
\bar{\psi}_2 \rangle = \frac{1}{2} \sum_{ii'} P_{i}^L P_{i'}^L |
 B_{ii'} \rangle \nonumber \\
+ \sum_{i} \left[ P_{i}^L {\hat
Y}_{i}^{\dag}- \bar{P}_{i}^L {\hat X}^{\dag}_{i}\right]
|\bar{\psi}_1 \rangle. \label{bar2hole}
\end{eqnarray}
Recalling that $|B_{ii'} \rangle$, Eq.\ (\ref{B}), is the
interacting two--exciton state, we see that the first term on the
rhs of the above equation describes the  $X$--$X$ interaction
effects similar to the undoped case.
\cite{ost-98,shah00,chernyak98,kner-97-99,fromer-00} The
second term describes the scattering of $X_i$ 
 with the carriers in the \{1-$h$/2DEG*\} state
$|\bar{\psi}_1\rangle$.

Finally, we turn to the 0-$h$ state. In Eq.\ (\ref{0hole}) we 
split this state into the contribution of the ground state $|0\rangle$,
with amplitude $\langle 0 | \psi \rangle= \langle 0 | \psi_0
\rangle$, and the \{0-$h$/2DEG*\} contribution $|\psi_0^{int}
\rangle$. The latter 2DEG* contribution 
is generated by the two--photon processes of excitation
and de-excitation of the system  accompanied
by the scattering of 2DEG excitations, 
and is further decomposed into two  parts. The
first part, $-\sum_{i} P_{i}^{L*}{\hat X}_{i}
|\bar{\psi}_1\rangle$, describes the de-excitation, after time
$t$,  of $X_i$ from the 
\{1-$h$/2DEG*\} state $|\bar{\psi}_1\rangle$
without scattering with the $|\bar{\psi}_1\rangle$
carriers. The latter scattering, as well as the
time evolution of the  2DEG excitations created via  second--order
processes analogous to the ones that lead to the inelastic Raman
scattering signal \cite{raman}, are described by the second part,
$| \bar{\psi}_0 \rangle$.

The 0--$h$ state $|\psi_0\rangle$ 
contributes to the third--order 
nonlinear response to second order in the optical field. By substituting
Eq.\ (\ref{0hole}) into Eq.\ (\ref{psi0}) and using Eqs.\
(\ref{Yop}), (\ref{linpol}), 
and (\ref{bar1hole}), we obtain the
equation of motion 
\begin{eqnarray}
i \partial_t | \bar{\psi}_0 \rangle -
 H | \bar{\psi}_0 \rangle
\, = \, \sum_{ii'} P_{i}^{L*} P_{i'}^L \, \hat{X}_i \, |Y_{i'} \rangle 
\nonumber  \\
+
\sum_{i}
  \left[ P_{i}^{L*} {\hat Y}_{i} -
{\bar P}^{L*}_{i} {\hat X}_{i} \right] | \bar{\psi}_1 \rangle 
-\sum_{ii'} P_{i}^{L*} \bar{P}^L_{i'} \, \hat{X}_i \, 
 | X_{i'}\rangle \nonumber \\
- \mu {\cal E}^* \ \sum_{ii'} N_i^{1/2} P_{i'}^L \, 
\left( [\hat{X}_i, \hat{X}^\dag_{i'}]
 - \delta_{ii'} \right) | 0 \rangle. \ \ 
 \label{bar0hole}
\end{eqnarray}
The first term in Eq.\ (\ref{bar0hole})
describes the  photo-excitation of the 2DEG via 
the second--order process where the exciton $X_{i'}$, photo-excited 
with amplitude $P_{i'}^L$,
scatters with the 2DEG into the state
$|Y_{i'}\rangle$, and then the exciton $X_i$ is deexcited 
with amplitude $P_{i}^L$. The above process leaves the system 
in a 2DEG$^*$ state. It is analogous to the photoexcitation 
of coherent phonons in  undoped semiconductors,
and dominates the 
inelastic Raman
scattering spectra of the 2DEG. 
\cite{raman}
The second term on the rhs of Eq.\
(\ref{bar0hole}) describes the scattering of $X_i$ with the
carriers in $|\bar{\psi}_1\rangle$
during its de--excitation. The rest of the terms 
describe the possibility to create  2DEG  excitations 
by photoexciting an exciton whose hole 
then recombines with a 2DEG electron.

\section{Nonlinear Polarization equation of motion}

 \label{NLPeom}

In this section we 
derive the equation of motion of the third--order nonlinear
polarization $P_i(t)$ that determines 
the FWM and 
pump--probe nonlinear optical signal, 
and separate out the factorizable  from the   
correlated contributions.
When we
discuss the physical meaning of the different terms
we focus on the FWM.

By taking the time derivative of Eq.\ (\ref{Pi}) and using the
definition of the operator ${\hat Y}_i$, Eq.\ (\ref{Yop}),
we obtain that
\begin{eqnarray}
 i \partial_t P_i(t) - \Omega_i P_i(t) +\sum_{i' \ne
i} V_{ii'} P_{i'}(t) = \nonumber 
\\
-\mu {\cal E}(t) \ \sum_{i'} \, N_{i'}^{1/2} \, \langle 
\psi | [\hat{X}_i, \hat{X}^\dag_{i'}] | \psi \rangle 
 + 
\langle \psi | {\hat
Y}_i | \psi \rangle.  \label{eom1}
\end{eqnarray}
The first term on the rhs of the above equation 
describes the Pauli blocking effects,
which only lead to positive time delay FWM signal
and are determined by the density of the LLi photoexcited 
carriers (recall Eqs. (\ref{commut}) and (\ref{DN})).  
The second term describes the optical signal
generated by the interactions between 
the recombining exciton $X_i$ leading  to the 
coherent emission  and 
the photo-excited and 2DEG carriers. 
This interaction contribution dominates the 
FWM signal for negative time delays.\cite{chemla-01}
The above two source terms  can be obtained 
by considering their equations of motion, 
which leads to an infinite hierarchy
of equations of motion.  
Alternatively, one can first separate out the 2DEG$^*$ 
and the 
X--X and X--$|\bar{\psi}_1\rangle$
interaction contributions 
by using  the decompositions of the photo-excited state, 
Eqs.\ (\ref{1hole}), (\ref{2hole}), and (\ref{0hole}), 
and by retaining  contributions up to third order 
in the optical field. 
Using the  property $\langle 0 | {\hat X}_{i'} {\hat Y}_i
= \langle 0 | {\hat Y}_i  {\hat X}_{i'} + \langle 0 |B_{ii'} $ 
(Eq.\ (\ref{B})), 
the expansion Eq.\ (\ref {psiexpansion}), 
and some algebra  we then obtain that
\begin{eqnarray}
\langle \psi |{\hat Y}_i | \psi \rangle = 
 \sum_{i'} P_{i'}^{L*} \langle  B_{ii'} |\psi_2
\rangle 
+ \sum_{i'} P_{i'}^{L}  
  \langle M_{ii'} | \bar{\psi}_0 \rangle^{*}  \nonumber
\\
+ \frac{1}{2} \,  \sum_{i'j'} P_{i'}^L \, P_{j'}^L 
\langle \bar{\psi}_1 | [[{\hat Y}_i, {\hat X}^{\dag}_{i'}], {\hat
X}^{\dag}_{j'}] | 0 \rangle \nonumber \\
+\sum_{i'} P_{i'}^{L} \, \langle
\bar{\psi}_1 | [{\hat Y}_i, {\hat X}^\dag_{i'}] | \bar{\psi}_1
\rangle
+ \bar{P}_i,
\label{interactions} 
\end{eqnarray}
where we introduced the state 
\begin{equation}
\label{Mstate} | M_{ii'} \rangle = {\hat Y}_{i} |X_{i'}
 \rangle.
\end{equation} 
Noting that $\langle 0 | M_{ii'} \rangle =0$, 
we see that the above state describes an excited 2DEG 
configuration 
with full valence band. 
The first term in Eq.\ (\ref{interactions}) describes 
the $X$--$X$ interactions. Its equation 
of motion can be obtained
by projecting the state  
$\langle  B_{ii'} |$ to Eq. (\ref{psi2}). 
In many cases it is useful to decompose the above  contribution 
into HF and correlated X--X interaction contributions
by using Eq.(\ref{2hole}): 
\begin{eqnarray} 
\langle  B_{ii'} |\psi_2
\rangle = 
\frac{1}{2} \,
\sum_{j'j} \,  \langle  B_{ii'} | X_{j}  X_{j'} \rangle \,
P_{j}^L \ P_{j'}^L \nonumber \\
+ \sum_{j} \, P_j^L  \langle  B_{ii'} | X^{\dag}_{j} 
| \bar{\psi}_1 \rangle \,
+{\cal B}_{ii'}.\label{2X} 
\end{eqnarray} 
The first term describes  the 
HF X--X interactions, 
familiar from the undoped case. \cite{ost-98}
The second term 
comes from the  exchange
process where
the first optical transition creates 
the $\{1-$h$/2DEG*\}$ state $|\bar{\psi}_1\rangle$
and the second transition
excites an $X_j$ $e$--$h$ pair 
while returning the 
conduction electrons to their ground state.
The above 
process  results 
in two $e$--$h$ pairs, which scatter with each other 
while the 2DEG is at rest. 
Subsequently, one of the above 
pairs, $i'$, is de--excited by the optical field,
while the remaining pair, $i$,  recombines and leads to 
the coherent emission. 
The last term in Eq. (\ref{2X}) 
describes 
biexciton and X--X scattering correlations.  
Similar to the undoped case,
 \cite{ost-98,shah00} such effects are
 characterized 
by the  amplitude of the 
correlated 2--$h$ photo-excited state,
\begin{equation}
\label{XXcorr}
{\cal B}_{ii'}(t) = \langle B_{ii'}  | {\bar
\psi}_2 \rangle.
\end{equation}
\begin{figure}[t]
\begin{center}
\includegraphics*[width = 8.5cm,height=4cm]{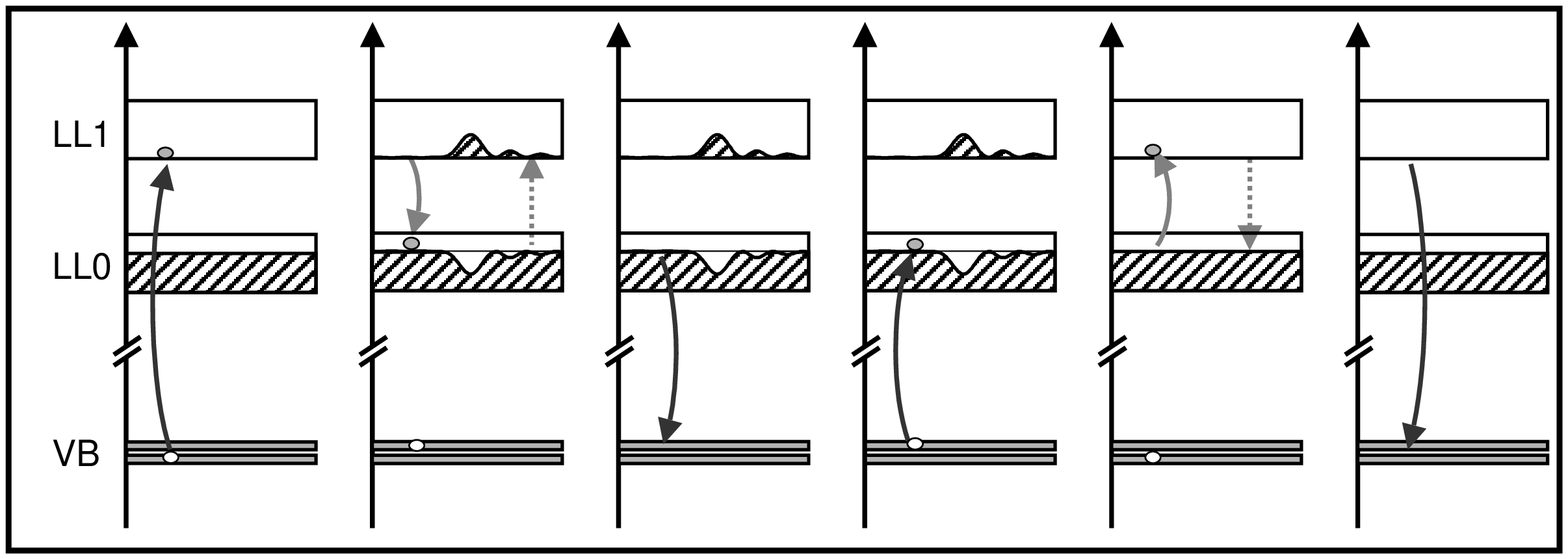}
\caption{ 
Dominant resonant process 
determining the  
MP correlation contribution to the FWM signal.
The first three panels show the Stokes Raman scattering 
that creates the MP, while the other three panels 
show the reverse  process that returns the system 
to the ground state 
}
\label{MPC}
\end{center}
\end{figure}
The  effects due to the propagation in time 
of the intermediate 2DEG excitations,  
photoexcited via the two--photon process
in Fig. \ref{FWM-process}(a), 
are described 
by the 
amplitude 
\begin{equation}
\label{C} {\cal M}_{ii'}(t) = 
\langle  M_{ii'} |
 \bar{\psi}_0 \rangle.
\end{equation}
Such time propagation leads to non--Markovian effects. 
In the case of particular interest here, 
the corresponding resonant 
contribution to the FWM signal
is due to the  nonlinear optical process 
shown  
in Fig. \ref{MPC}.
The X
photo-excited by the first optical 
transition  decays into $Y$
excitation. The 
$e$-$h$ pair in this X+MP state then recombines,
 leading to coherent
emission, and leaves the system in a MP state.
This MP propagates in time and then scatters 
with the second photo-excited $X$ into an $X$ state subsequently
annihilated by the  optical field. It is interesting to note the
similarity of this process and the familiar one of coherent
antiStokes Raman scattering \cite{levenson-82} that, however,
involves phonons. 

The second line in Eq.\ (\ref{interactions}) 
describes a
shake-up of the 2DEG during the exciton recombination
that gives  
the coherent emission. In particular, the photo-excitation of
two excitons, $X_{i'}$ and $X_{j'}$,
is followed by the recombination of one
them assisted by the shake--up of a 2DEG excitation. The above
process leaves the system in a \{1-$h$/2DEG*\} state, 
which is then annihilated by the optical  field.
To interpret the first term on the third line 
of Eq. (\ref{interactions}),
we
note that
the HF XX interaction 
can  be thought of as arising from the scattering of the 
polarization with the coherent density. \cite{staff90}
Similarly, this
term describes the scattering 
of the polarization with the incoherent density
of photoexcited carriers in the \{1-$h$/2DEG*\} state
$| \bar{\psi}_1\rangle$. 

Finally, the last term on the rhs of Eq.\ (\ref{interactions}) is
the correlated contribution 
\begin{eqnarray}
\bar{P}_i = \langle \psi_0 | 0 \rangle \,  \langle Y_i | \psi_1
\rangle + \sum_j P_j^{L*} \langle Y_i X_j  | \psi_2 \rangle \nonumber
\\
+
\langle {\bar \psi}_0 | {\hat Y}_i | {\bar \psi}_1 \rangle+
\langle {\bar \psi}_1 | {\hat Y}_i | {\bar \psi}_2 \rangle.
\label{P_deph}
\end{eqnarray}
The first two terms of Eq.\ (\ref{P_deph})
describe the dephasing of the $X$ and 2-$X$ amplitudes that
determine the third--order nonlinear polarization, 
while the last two terms describe the dephasing of the incoherent
contribution to the nonlinear polarization.
Note that, by linearizing the above equation, we recover the
correlation function $\bar{P}_i^L$, Eq. (\ref{barpol}), that 
describes the dephasing of the linear polarization  $P_i^L$.

Using the above results 
we obtain the following equation of motion for the
third--order nonlinear  polarization:
\begin{eqnarray}
i \partial_t P_i -   \Omega_i P_i + \sum_{i' \ne i} V_{ii'}
P_{i'} -  \bar{P}_i \hspace{2cm} \nonumber
\\ 
= \mu {\cal E} \ \sum_{i'} \, N_{i'}^{1/2} 
 \, \langle \psi_1 | \left(\delta_{ii'} 
-[\hat{X}_i, \hat{X}^\dag_{i'}] \right) | \psi_1 \rangle 
\nonumber \\
+ 
\frac{1}{2} \,\sum_{jj'i'} \,
  \langle  B_{ij} | X_{i'}  X_{j'} \rangle \, 
P_{i'}^L \ P_{j'}^L P_{j}^{L*} 
+\sum_{i'} {\cal B}_{ii'} P_{i'}^{L*}
\nonumber \\
+ \sum_{i'}  P_{i'}^{L} {\cal M}_{ii'}^*
\nonumber \\
+  \frac{1}{2} \sum_{i'j'} P_{i'}^L \, P_{j'}^L 
\langle 0 | [\hat{X}_{j'},[{\hat X}_{i'},{\hat Y}_i^\dag ]]| 
 \bar{\psi}_1 \rangle^{*}
\nonumber \\
+ \sum_{i'}  P_{i'}^{L}
\langle \bar{\psi}_1 | [{\hat
Y}_i, {\hat X}^\dag_{i'}]
 | \bar{\psi}_1 \rangle 
+ \sum_{ji'} \, P_{i'}^L  P_{j}^{L*} 
\, \langle  B_{ij} | X^{\dag}_{i'} 
| \bar{\psi}_1 \rangle. 
\ \ \ \label{eom}
\end{eqnarray}
In the above equation we have separated out the 
source terms into (i) a coherent part, determined by 
Pauli blocking effects, 
 HF XX interactions, 
and the propagation 
in time of the intermediate interacting XX and 2DEG$^*$ 
states (first three lines on the rhs), 
and (ii) an incoherent part, determined by the 
 \{1-$h$/2DEG*\} photoexcited state $|\bar{\psi}_1 \rangle$
(last two lines on the rhs)
and the correlated nonlinear contribution 
$\bar{P}_i$.
It is worth noting in the above equation  that
the terms 
 proportional to the 
polarization $P_{i'}^L$ 
describe a time--dependent 
photo--induced renormalization 
of the $X_i$ energy and dephasing
( $i'=i$),  
and of the coupling $V_{ii'}$ 
of the $X_{i}$ and $X_{i'}$ 
states ( $i' \ne i$).
If the 
\{1-$h$/2DEG*\} and 
$|B_{ii'}\rangle$ 
excitations decay rapidly, 
while  the 2DEG excitations are long lived,  
then the non--Markovian LL coupling 
and dephasing effects are dominated 
by the correlation function ${\cal M}_{ii'}(t)$.

As demonstrated by the above equation, 
the Coulomb correlations lead to 
new contributions to the  nonlinear polarization, 
determined 
by  many--particle correlation functions. 
In the next section 
we  turn to the problem of solving for the  
correlation functions on the rhs of Eq. (\ref{eom}),
 and address the issue 
of dephasing in a strongly correlated electronic system.

\section{Dephasing and correlation processes}
\label{dephasing}

The  equations of motion 
for the correlation functions 
that enter in Eq. (\ref{eom}) 
may be obtained after  introducing a basis suitable 
for describing the 
1$h$, 2$h$, and 0$h$  states.
In  strongly correlated systems, an expansion 
in terms of the pair excitations 
of a non--interacting many--electron  may not be convenient. In
general we must introduce a basis of strongly 
correlated states that 
already incorporate the ground state correlations. 
 The choice of such a basis 
depends  on the 
ground state 
and on the most important  excitations 
for the experimental parameters of interest 
(e.g. the filling factor, the central excitation frequency, 
the polarization of the optical field, etc).
 
Our goal in the rest  of this paper 
is  to identify the dominant features in  the 
FWM spectrum 
at the 
magnetoexciton energies that come from
the interactions and time propagation 
of the MP collective modes. 
Analogous questions regarding the role of X--X interactions in undoped 
semiconductors were first addressed by using average polarization models.
\cite{chemla-99,schaf96}
As we show in the next section, 
a generalized average polarization model
can be extracted 
from the theory developed in the previous sections
after introducing  a basis of Lanczos strongly 
correlated states. \cite{haydock}
This model explains the main qualitative  features 
observed in recent FWM experiments. \cite{from-02} 

We  start by considering 
a basis for the 
1--$h$ state $| \bar{\psi}_1\rangle$. 
Noting the analogy with X--phonon interactions 
in the undoped system, discussed in the introduction, 
we would like to consider a
basis  that consists of  products of 
$e$--$h$ pair and  MP 
wavefunctions. In the undoped system, 
such 
 states 
have the form $\hat{e}^\dag_{{\bf k-q}} \hat{h}_{{\bf k}}^\dag
a^\dag_{{\bf q}} | 0 \rangle$, 
where $a^\dag_{{\bf q}}$ creates the phonon 
state, $\hat{e}^\dag_{{\bf k-q}}
\hat{h}_{{\bf k}}^\dag$ 
creates the two--particle $e$--$h$ pair wavefunction, 
and  $|0\rangle$ is 
annihilated by all the 
$\hat{e}$ and $\hat{h}$ operators.
\cite{axt98,axt96,schilp-94}
In our case however, the ground state $|0\rangle$ may
contain a strongly correlated electron gas, 
while, unlike for phonons, 
the MP creation operators are made of electrons.  
Thus we must 
use a basis of strongly correlated 
states that is made out of
electrons.

A basis set useful for calculating 
 Green functions 
for tight binding
and Hubbard Hamiltonians is the 
Lanczos basis.\cite{haydock}
Such correlated states can be used to obtain exact solutions 
in the case of small 
systems \cite{chak-book}, but also to approximately describe 
continuum resonances in large systems, e.g. the 
Fano resonances 
in the absorption spectrum of semiconductor
superlattices. 
\cite{chu} 
Each new basis state is obtained 
by acting with the Hamiltonian $H$ on the previous 
state, and then
orthogonalizing the result with respect to all existing basis
states. \cite{haydock}
This procedure is similar to Eq.\ (\ref{HonX}) 
that introduced the states $| Y_i
\rangle$, and led us
to the parameters $\Omega_i$ and $V_{ii'}$, 
 Eqs.\ (\ref{Xen}) and (\ref{V}). A new basis  state
$|Z_i\rangle= {\hat Z}_i^\dag |0\rangle$ is now constructed from
the relation
\begin{equation}
H |Y_i \rangle = \bar{\Omega}_i |Y_i \rangle +  \sum_{i'} W_{i'i}
|X_{i'} \rangle + |Z_i \rangle , \label{HonY}
\end{equation}
where
\begin{equation}
\label{Yen} \bar{\Omega}_i =\frac{\langle Y_i |  H | Y_i \rangle}{
\langle Y_{i}|Y_{i}\rangle}
\end{equation}
is the average energy of the four-particle excitation
$|Y_i\rangle$,
\begin{equation}
\label{W} W_{i'i} = \langle X_{i'}| H  | Y_i \rangle
\end{equation}
gives the probability amplitude that $Y_i$ scatters into $X_{i'}$,
and we introduced the operator
\begin{equation}
\label{Zop} {\hat Z}_i = [{\hat Y}_i,H] - \bar{\Omega}_i {\hat
Y}_i - \sum_{i'} W_{ii'} {\hat X}_{i'}. 
\end{equation}
Using Eqs.\ (\ref{HonY}), (\ref{Yen}), and (\ref{W}), as well as
the orthogonality $\langle X_{j}| Y_i \rangle=0$,  one can see
that the state $| Z_i \rangle$ is orthogonal to all the states $|
X_j \rangle$, $j=0,1,\cdots$, and to $|Y_i \rangle$. Therefore, it
is a linear combination of all the 2DEG* states into which $|Y_i
\rangle$ can scatter.
Additional basis states can be 
constructed 
by applying the above orthogonalization 
procedure to the state  $H | Z_i \rangle$. 
 We note here  that the states $|X_j \rangle, 
| Y_i \rangle, 
| Z_i \rangle, \cdots$ 
do not correspond to an 
 expansion in terms of excitations of 
a noninteracting many--electron  state, since 
they 
are  obtained 
by the action of the operators $\hat{X}_j^\dag, \hat{Y}_i^\dag,
\hat{Z}_i^\dag, \cdots$   on the
ground eigenstate $|0>$ of the many--body Hamiltonian 
$H$.

By using Eq.\ (\ref{HonX}) and the orthogonality $\langle X_j |
Y_i \rangle=0$ we obtain the useful relation
\begin{equation}
 \label{W1}
W_{i'i} = \langle Y_{i'}| Y_i \rangle = W^{*}_{ii'}.
\end{equation}
Note that  $\langle Y_{i'}| Y_i \rangle \ne 0$, and we may also
have that $\langle Y_{i'} |Z_i \rangle \ne 0$ for 
$i' \ne i$. If
this is the  case we  need to orthogonalize the independent
states $| Y_{i'} \rangle$, and then subtract a linear combination
of the latter from $|Z_i \rangle$ in Eq.\ (\ref{HonY}) so that all
the  $Z$ and $Y$  states become orthogonal. However, in the
electron--hole symmetric limit of the 2DEG system,  $| Y_{i}
\rangle$ is the same state for all $i$  
when only LL0 and LL1 contribute (see Appendix \ref{symm}),
and thus the latter
procedure is not needed.

Equations of motion for all correlation 
functions determined by the 
state $| \bar{\psi}_1\rangle$
can  be obtained after expanding in a basis set
of 
\{1-$h$/2DEG*\} states. Let us consider for example 
$\bar{P}_i^L$(t), Eq.\ (\ref{barpol}), which
describes the dephasing of the linear polarization $P_i^L$(t).
If we choose the Lanczos basis discussed above, 
we obtain after multiplying Eq.\ (\ref{bar1hole}) by $\langle Y_i|$ 
and using
Eqs.\ (\ref{HonY}), (\ref{Yen}), and (\ref{W})
the equation of motion 
\begin{equation}
\label{barpol_eom} i \partial_t \bar{P}_{i}^L =  \bar{\Omega}_{i}
\bar{P}_{i}^L + \sum_{i'} W_{ii'} P_{i'}^L + {\cal Z}_i^L
\end{equation}
where we introduced the correlation function ${\cal Z}_i^L=
\langle Z_i |  {\bar \psi}_1 \rangle$ 
whose equation 
of motion can be obtained in a similar way
as that of $\bar{P}_i^L$.  

It is important to note that the dephasing of the optical
polarization obtained as above is non--Markovian.
Indeed, after solving  Eqs.\ (\ref{linpol}) and
(\ref{barpol_eom}) by Fourier transform
we obtain that: 
\begin{eqnarray}
 \left[\omega - \Omega_i(\omega) \right]
P_{i}^L(\omega) + \sum_{i' \ne i} V_{ii'}(\omega) P_{i'}^L(\omega)
\nonumber  \\
=- \mu {\cal E}(\omega) N_i^{1/2} + \frac{{\cal Z}_i^L(\omega)}{\omega 
- \bar{\Omega}_i}, \label{fourpol}
\end{eqnarray}
where the X energy $\Omega_i(\omega)$ 
and the coupling between the  X states
$V_{ii'}(\omega)$ 
include 
frequency--dependent   self--energy corrections due
to the X--2DEG scattering,
\begin{equation}
\label{EV_omega} \Omega_i(\omega) = \Omega_i + \frac{W_{ii}}{
\omega - \bar{\Omega}_i}
\ , \ 
V_{ii'}(\omega) = V_{ii'} + \frac{W_{ii'}}{\omega
- \bar{\Omega}_i}.
\end{equation}
Additional self--energy corrections arise from 
${\cal Z}^L$, discussed in the next section.  
The  frequency--dependence of the  above X energies
and coupling constants is a manifestation of the non--Markovian
behavior of the system. This arises because part of the optical
excitation is temporarily stored in the shake--up excitations
described by $\bar{P}^L_i$.

Using the recursive method we can also construct a basis for the
2--$h$ and 0--$h$ states, which we can then use to calculate the
correlation functions 
determined by the states $|\bar{\psi}_2 \rangle$ 
and  $|\bar{\psi}_0 \rangle$.
We start with the 2--$h$ state 
$|B_{ii'}\rangle$ that determines the 
XX correlation function 
${\cal B}_{ii'}$, and introduce the Lanczos 
state $|\bar{B}_{ii'} \rangle$ as follows: 
\begin{equation}
\label{HonB} H |B_{ii'}\rangle = \Omega_{ii'}^B |{\cal
B}_{ii'}\rangle + |\bar{B}_{ii'}\rangle  \ , \
\Omega_{ii'}^B = \frac{ \langle B_{ii'} | H |B_{ii'}\rangle}{
 \langle B_{ii'}  | B_{ii'}\rangle},
\end{equation}
where $\Omega_{ii'}^B$ is the average energy of the interacting
2-$X$ state $|B_{ii'} \rangle$. The state 
$|\bar{B}_{ii'}\rangle$, $\langle \bar{ B}_{ii'}| B_{ii'}\rangle=0$,
 is a 
linear combination of all the 2-$X$  states into which $|
B_{ii'}\rangle$ can scatter.
By projecting the state $\langle B_{ii'}|$ 
to Eq. (\ref{bar2hole})
and using 
 Eq.\ (\ref{HonB}),
we then obtain
from Eq. (\ref{XXcorr}) 
 the equation of
motion
\begin{eqnarray}
 i \partial_t {\cal B}_{ii'} - \Omega^B_{ii'}  {\cal
B}_{ii'} = \frac{1}{2}  \sum_{j'j} \, \langle B_{ii'} |
B_{jj'} \rangle \,  P_{j'}^L P_{j}^L \  \  \ \nonumber  \\
+ \sum_{j} P_{j}^L
\langle  B_{ii'} |  {\hat Y}_{j}^{\dag} |\bar{\psi}_1 \rangle
- \sum_j \bar{P}_{j}^L
\langle  B_{ii'} | 
{\hat X}_{j}^{\dag}|\bar{\psi}_1 \rangle
+ \bar{{\cal B}}_{ii'},\ \  \label{Beom}
\end{eqnarray}
where $ \bar{{\cal
B}}_{ii'}(t) = \langle \bar{ B}_{ii'} | \bar{\psi}_2 \rangle$.
The above equation 
describes the time evolution of the ``intermediate'' 
2-$X$ state $| B_{ii'}\rangle$, which is created by the
$X$--$X$ interactions.
In the case of undoped QW magnetoexcitons,   
${\cal B}_{ii'}(t)$ 
corresponds to $F(t)$ of Ref. \onlinecite{shah00}.

One should note here the similarity of Eq.\ (\ref{Beom}) and the
average polarization model that has been successful in describing
the $X$-$X$ correlations and biexciton effects in undoped
semiconductors. \cite{chemla-99,shah00,schaf96} 
 This model includes the 
XX self--energy effects due to 
the higher Lanczos states 
$|\bar{B}_{ii'} \rangle,\cdots$ 
via  a phenomenological dephasing rate. 
The validity of such a model 
in the case of undoped QW magnetoexcitons 
was analyzed in Ref. \onlinecite{shah00}.

Similar to   ${\cal B}_{ii'}$,
the correlation function ${\cal M}_{ii'}$ describes the time
evolution of  the ``intermediate'' 
photo-excited 2DEG$^*$ state $| M_{ii'} \rangle$.
Using the Lanczos method we obtain that
\begin{equation}
\label{HonM} H | M_{ii'}\rangle = \Omega_{ii'}^M |
M_{ii'}\rangle + |\bar{ M}_{ii'}\rangle \ , \
\Omega_{ii'}^M = \frac{\langle  M_{ii'} | H |
M_{ii'}\rangle}{
 \langle  M_{ii'}  | M_{ii'}\rangle},
\end{equation}
where $\Omega_{ii'}^M$ is the average MP energy, and the state
$|\bar{ M}_{ii'}\rangle$, $\langle \bar{M}_{ii'}| 
M_{ii'}\rangle=0$, is a linear combination 
of all the states into which  $| M_{ii'} \rangle$ 
can scatter. 
We then obtain, 
after projecting  $\langle  M_{ii'}|$
on Eq.\ (\ref{bar0hole}), the equation of motion 
\begin{eqnarray}
i \partial_t {\cal M}_{ii'} - \Omega^{M}_{ii'} 
{\cal M}_{ii'}  - \bar{{\cal M}}_{ii'}
=  \sum_{jj'} \langle M_{ii'} |  \hat{X}_j | Y_{j'} 
 \rangle \, P_j^{L*} P_{j'}^L
 \nonumber \\
+ \sum_j P_j^{L*} 
\left[
 \langle M_{ii'} |  {\hat Y}_j | {\bar \psi}_1 \rangle
- \sum_{j'} \, 
\langle M_{ii'} | \hat{X}_j\, | X_{j'}\rangle \
\bar{P}^L_{j'} \,
\right] \nonumber \\
- \sum_j {\bar P}^{L*}_j 
\langle  M_{ii'} | {\hat X}_j | {\bar \psi}_1 \rangle 
\nonumber \\
+ \mu {\cal E}^* \sum_{j'j} N_j^{1/2} \, P_{j'}^L \, 
\langle M_{ii'}| \left( \delta_{jj'} - [\hat{X_j}, \hat{X}^\dag_{j'}]
\right) |0 \rangle, \ \ \ \
\label{Ceom}
\end{eqnarray}
where $ \bar{{\cal M}}_{ii'}= \langle \bar{M}_{ii'}| \bar{\psi}_0 
\rangle$ describes the 
dephasing of ${\cal M}_{ii'}$. 
In the case of the 2DEG, 
the  single mode approximation \cite{chak-book}  
suggests that the latter dephasing
can be treated to first approximation 
by introducing a phenomenological dephasing rate.

The  remaining step is the calculation of the correlated contribution 
$\bar{P}_i$, Eq.\ (\ref{P_deph}). 
The equation of motion 
for the first two terms of 
Eq. (\ref{P_deph}) can be easily obtained 
from Eqs. (\ref{psi1}) 
and (\ref{psi2}) after using 
Eq. (\ref{HonY}) 
and the property 
\begin{eqnarray} 
\langle Y_i X_j | H = 
(\bar{\Omega}_i + \Omega_j ) 
\langle Y_i X_j |
- \sum_{j' \neq j} V_{jj'} 
\langle Y_i X_{j'} |
\\+ \sum_{i'} W_{ii'} \langle X_{i'} X_j |
+ \langle Y_i|  \hat{Y}_j 
+ \langle Z_i| \hat{ X}_j,\nonumber 
\end{eqnarray}
obtained by calculating the commutator 
$[\hat{Y}_i \hat{X}_j,H]$ using 
Eqs. (\ref{Yop}) 
and (\ref{HonY}).
The equations of motion for the last two terms 
in Eq. (\ref{P_deph}) 
can be obtained
from Eqs. (\ref{bar0hole}),  (\ref{bar1hole}),  
and (\ref{bar2hole}) by using 
Eq. (\ref{Zop}) and the basis of choice.
We thus obtain 
the equation of motion 
\begin{eqnarray}
i \partial_t \bar{P}_i - \bar{\Omega}_i \bar{P}_i - {\cal Z}_i =
\sum_{i'} W_{ii'} P_{i'} + {\cal Q}_{Y_i},
\label{Pbar}
\end{eqnarray}
where the correlated contribution 
\begin{eqnarray}
 {\cal Z}_i = \langle \psi_0 | 0 \rangle \,  \langle Z_i
| \psi_1 \rangle + \sum_j P_j^{L*} \langle  Z_i| {\hat X}_j |
\psi_2 \rangle \ \ \ \nonumber  \\
+\langle \bar{\psi}_0 | {\hat Z}_i | \bar{\psi}_1
\rangle+ \langle {\bar \psi}_1 | {\hat Z}_i |  \bar{\psi}_2
\rangle \label{calZ}
\end{eqnarray}
has the same structure as $\bar{P}_i$ (with 
the difference $\hat{Y}_i \rightarrow \hat{Z}_i$) and 
describes  the dephasing of $\bar{P}_i$. 
The factorizable  contribution
${\cal Q}_{Y_i}$  
describes photo--induced nonlinear corrections  to the 
dephasing and energy of $\bar{P}_i$, and to the scattering amplitudes 
$W_{ii'}$. 
The equation of motion 
for ${\cal Z}_i$, which  
to first order in the
optical field coincides with ${\cal Z}_i^L$,
has a form analogous
to that of $\bar{P}_i$.

One should note here that the correlation function 
$\bar{P}_i$ can be decomposed further 
 in the case of systems 
where the X--X interaction 
contribution
to the operator $\hat{Y_i}$
  can be separated out.
This is possible for example 
in undoped semiconductors \cite{axt96,axt98}, 
where the 
operator $\hat{Y_i}$  can be decomposed into a part 
that is independent of the phonon variables, which  
describes  the $X$-$X$ Coulomb
interactions, and a part that describes the phonon
creation/annihilation processes.
The former $X$-$X$ contribution 
comes from the last term 
in Eq.\ (\ref{P_deph}),  and corresponds 
to  the correlation function $\bar{Z}$ of
Refs.\ \onlinecite{axt96,axt98} that mainly contributes to the
six--wave--mixing spectra. \cite{bolton-00-01}
The above distinction between the 
interaction processes is possible 
in systems 
where the creation operators of the 
ground state excitations 
of interest 
(phonons, MPs, magnons, $\cdots$) 
commute with the electronic operators 
that describe the photoexcited carriers.

\section{Generalized Average Polarization Model}

\label{GPA} 

In this section we present an example 
of how the theoretical 
framework 
developed so far can be used to describe the 
nonlinear optical dynamics 
of the 2DEG in a high magnetic field.
We consider the case 
where only the first two LLs
are photo-excited, 
so we retain in our calculations only the LL0 and
LL1 magnetoexcitons.
We focus on  filling factors close to 
$\nu=1$,  where 
the ground state 
2DEG  populates spin--$\uparrow$ LL0 states, \cite{ferro,aif-96}  
and  on photoexcitation with 
$\sigma_{+}$ circularly polarized light, 
which excites 
spin--$\downarrow$ electrons.\cite{from-02} 
The above conditions apply to the 
experiment 
of Ref. \onlinecite{from-02} that  we wish to interpret. 

 The electron--hole symmetry
of the ideal 2D system, 
analyzed in Appendix \ref{symm}, 
relates the correlation functions 
and interaction parameters with different 
LL indices that enter in the equations of motion.
For  example, 
in Appendix \ref{symm} we  derive 
the symmetry property
\begin{equation}
\label{Y} 
\sqrt{1 - \nu_1} \ \hat{Y}_1 = 
- \sqrt{1 - \nu_0} \ \hat{Y}_0 = \hat{Y},
\end{equation} 
where $\hat{Y}$ is determined  by 
Eq. (\ref{Y-symm}).
The above symmetry relation can be used to
reduce the number of independent 
variables.
For example, 
from Eqs. (\ref{Yen}), (\ref{W1}), and 
(\ref{P_deph})
we obtain that $\bar{\Omega}_i = \bar{\Omega},$
$W_{10} = W_{01}$, 
\begin{eqnarray} \
\label{W-symm} 
(1 - \nu_i) \  W_{ii} =
-\sqrt{(1 - \nu_0) (1 - \nu_1)} \ W_{01}
=W = \langle Y | Y \rangle,  \ \  \  \\
\label{Pbar-symm} 
\sqrt{1 - \nu_1} \ \bar{P}_1(t)  = - \sqrt{1 - \nu_0} 
\bar{P}_0(t) = W \bar{P}(t). \ \ \ \ 
\end{eqnarray}
where $i=0,1$,
It is then  convenient to make the 
transformation 
\begin{equation} 
\label{p-transf} 
P_i \rightarrow  P_i
\sqrt{1 - \nu_i},
\end{equation} 
and redefine for simplicity 
\begin{equation}
\label{V-transf}
V_{ii'} 
\rightarrow 
V_{ii'}\sqrt{(1 - \nu_i)
(1 - \nu_{i'})} \ , \ 
W \rightarrow 
W (1 - \nu_0)
(1 - \nu_{1}).
\end{equation} 
Using the above relations 
we  obtain 
from Eq. (\ref{linpol}) 
the following equations of motion for the  
linear polarizations:
\begin{eqnarray}
i \partial_t P_{0}^{L}
 = (\Omega_0 - i \Gamma_0)  P_{0}^{L}  \nonumber  \\
-  \left[ \mu {\cal E}(t)   
+  V_{01} ( 1- \nu_1) P_{1}^{L}+ 
W (1 - \nu_1)
 \bar{P}^{L} 
\right],\label{linpol-symm-0} 
\\ 
i \partial_t P_{1}^{L}
 = (\Omega_1 - i \Gamma_1)  P_{1}^{L} \nonumber  \\
- \left[ \mu {\cal E}(t)  
+  V_{10} (1 - \nu_0) P_{0}^{L}
- W (1 - \nu_0)
\bar{P}^{L} 
\right].\label{linpol-symm-1} 
\end{eqnarray}
The above equations have the form of two coupled two--level 
systems, corresponding to the LL0 and LL1 magnetoexcitons. 
This form is due to the
zero--dimensional confinement induced by the 
QW potential and the magnetic field, 
which leads to the novel 2DEG properties.\cite{chak-book,QHE2}  
The Coulomb interactions 
renormalize the 
 Rabi energy $\mu {\cal E}$ 
by a mean (local) field correction 
proportional to the polarization  (analogous to the undoped system
\cite{staff90}), 
and by a 2DEG shake--up correction proportional 
to $\bar{P}^L$.

 As demonstrated by 
Eqs. (\ref{linpol-symm-0}) and  
(\ref{linpol-symm-1}),
the  polarization dephasing 
is determined, in addition 
to the phonon--induced dephasing rates
$\Gamma_i$, by 
the correlation function  
$\bar{P}^L$. 
For weak  $\Gamma_i$, 
$\bar{P}^L$ dominates.  
In the absence of magnetic field, 
$\bar{P}^L$ describes the shake--up of FS pair excitations, 
and leads to a
non--Markovian dephasing due to the 
 non--perturbative $h$--FS 
interactions.\cite{per-00,prim-00}
For the experimental conditions of interest here, 
$\bar{P}^L$ originates  primarily  from 
the X scattering 
to the continuum of X+MP states 
composed of an X and a MP with opposite 
momenta. An analogy can be drawn between 
the above  X+MP states and the continuum 
of X+X scattering states in undoped semiconductors.
\cite{chemla-01,chemla-99,shah00}

We now turn to the dephasing of the X+MP 
correlation function $\bar{P}^L$. This X+MP dephasing 
originates from the Coulomb--induced coupling of the 
Lanczos  states 
$|Y\rangle, |Z\rangle, | Z' \rangle, \cdots$. 
We note that
${\cal Z}^L$  and   higher 
correlation functions 
do not couple to 
$P^L$, only to the amplitudes corresponding to the previous and the next Lanczos states. 
For example, the equation of motion for ${\cal Z}^L$ 
reads
\begin{equation} 
i \partial_t {\cal Z}^L = \bar{\Omega}_Z {\cal Z}^L 
+ W_{ZY} \bar{P}^L + \langle Z' | \bar{\psi}_1 \rangle,
\end{equation} 
where we defined 
\begin{equation} 
\label{Z-par} 
\bar{\Omega}_Z = 
\frac{ \langle Z | H | Z \rangle}{\langle Z | Z \rangle} 
\ , \ W_{ZY} = \frac{ \langle Z | Z \rangle}{\langle Y | Y \rangle}
= \frac{ \langle Z | H | Y \rangle}{W}.
\end{equation} 
The amplitudes of the higher Lanczos states 
satisfy similar equations of motion. 
After taking the Fourier transform and 
using the above 
symmetry properties and  some algebra we obtain that
\begin{equation} 
\bar{P}^L(\omega) 
=  \frac{P_1^L(\omega) - P_0^L(\omega)}{\omega - \bar{\Omega} 
+ i \gamma_Y
- W_{ZY} \Sigma_{Z}(\omega) }, 
\end{equation} 
where $\gamma_Y$  is the dephasing rate.
The dephasing of $\bar{P}^L$ is thus described by the 
self energy $\Sigma_Z$,
\begin{equation}
 \label{self}
\Sigma_{Z}(\omega) = \frac{1}{
\omega - \bar{\Omega}_Z + i \gamma_Z  - W_{Z'Z} 
\Sigma_{Z'}(\omega)}, 
\end{equation} 
where  $\Sigma_{Z'}$ is  given by 
Eq. (\ref{self})  with $Z' \rightarrow  Z''$.

The above equation can be used to obtain 
a continued fraction 
expansion for the self energy.
In the case of an $N$--electron system, 
such an expansion 
terminates after $N$ iterations.
To obtain true dephasing 
for finite $N$, 
 we must introduce the damping rates 
of the Lanczos states, due to the  neglected degrees of freedom
of the $N \rightarrow \infty$ system. 
The convergence 
of the above self--energy expansion 
becomes more rapid  with increasing damping rates.\cite{chu}    
In the QHE literature, numerical calculations 
of the N--electron spectral functions 
have been 
extrapolated to   the 
$N \rightarrow \infty$ limit.\cite{chak-book} 

Eq. (\ref{self}) can   be solved analytically 
when the dispersion in the 
energies and matrix elements  of the higher Lanczos states 
is small as compared  to 
the frequencies of interest, 
so that the self energy  
is approximately 
the same for all the higher Lanczos states.\cite{haydock}   
This may be the case for example if the momenta close to the magnetoroton 
minimum, $q \sim 1/l$,  dominate.\cite{raman}
In the case of QW magnetoexcitons in undoped semiconductors, 
the validity of such an approximation for the X--X 
self energy was discussed in Ref. \onlinecite{shah00}. 

A microscopic determination 
of $\Sigma_{Z}(\omega)$ is beyond the scope of this paper. 
Here we describe the X+MP scattering in a way 
similar  to  the average polarization 
model description of the X--X scattering.\cite{chemla-99}
In particular, we assume that $\Sigma_Z$  is 
a sufficiently smooth function of frequency 
in the  range of interest, in which case 
its frequency dependence can be neglected
to first approximation,  
In the case of 2D magnetoexcitons in undoped semiconductors, 
a similar
 approximation 
was shown to apply for strong 
X--X interactions.\cite{shah00} 
We thus obtain the equation 
\begin{equation}
\label{barpol-symm} 
i \partial_t \bar{P}^L = ( \bar{\Omega} - i \gamma) 
\bar{P}^L + P_1^L - P_0^L,
\end{equation}
where the values of the renormalized $Y$ state energy 
\begin{equation} 
\bar{\Omega}
= \frac{ \langle Y | H | Y \rangle}{\langle Y | Y \rangle} 
+  W_{ZY} \,  \rm{Re} \, \Sigma_Z(\bar{\Omega}) 
\end{equation} 
and  dephasing rate 
\begin{equation}
\gamma = \gamma_Y - W_{ZY} \  \rm{Im} \, \Sigma_Z(\bar{\Omega}) 
\end{equation} 
are estimated here by fitting to the 
experimental linear absorption spectrum.\cite{from-02}
The above approximation 
works best for sufficiently large $\gamma$. 
Due to the  contribution to the Y state 
of finite momentum  MPs 
and  $e$--$h$ pairs,
 we expect that $\bar{\Omega} > \Omega_1$, 
where $\Omega_1$, Eq. (\ref{Xen}), is the energy  of the zero momentum LL1 
magnetoexciton.\cite{raman}

We now turn to the nonlinear polarization, 
determined by the equation of motion 
Eq.(\ref{eom}). 
First we consider the X--X interaction contribution, 
described by the second line 
on the rhs of Eq.(\ref{eom}).
For strong 
damping of the 2X states $| B_{ii'} \rangle$, 
the non--Markovian X--X scattering 
effects
are suppressed\cite{shah00,ost-98}, and we only retain the 
HF X--X interaction contribution
(second term on the rhs of Eq.(\ref{eom})).
The X--X potentials
$\langle B_{ii'} | X_j X_{j'} \rangle$,   
with different LL indices $i$ and $i'$, 
are related to each other 
in the electron--hole symmetric limit 
due to the 
property 
\begin{equation} 
\label{Bsymm} 
(1 - \nu_i) | B_{ii}\rangle = 
-\sqrt{(1 - \nu_0) (1 - \nu_1) } | B_{10}\rangle,
\end{equation} 
where $i=0, 1$, that follows 
from Eq.(\ref{B-symm}).
Using the above relation and Eq.(\ref{p-transf}),
we express the HF XX interaction contribution in the form
\begin{eqnarray}
\label{XXpot}  
\sum_{jj'} \sqrt{( 1 - \nu_{j}) (1 - \nu_{j'})}
\, \langle B_{ii}| X_{j} X_{j'} \rangle \, P_{j}^L  
P_{j'}^L \, \left( P_{i}^{L*} - P_{i'}^{L*} \right),
\ \ \end{eqnarray} 
where $i' \ne i$ and the potential 
$\langle B_{ii}| X_{j} X_{j'} \rangle$ 
is evaluated in Appendix \ref{XX}.

The time--dependence 
of the incoherent source terms in the 
last two lines of Eq. (\ref{eom}) is determined 
by the 
\{1-$h$/2DEG*\}  state $| \bar{\psi}_1\rangle$. 
The corresponding correlation 
functions  dephase rapidly 
in the case of strong X+MP damping, 
unlike e.g.  the  correlation function 
${\cal M}$ that describes the time propagation 
of the long--lived MPs.
The same holds 
for the FWM contributions  due to the  photo--induced 
renormalizations of the X+MP correlation 
functions $\bar{P}$, ${\cal Z}, \cdots$, which are described by 
the source terms 
${\cal Q}_{Y}$,  ${\cal Q}_{Z}, \cdots$
in Eq. (\ref{Pbar}). The latter lead to 
an incoherent FWM contribution 
at the 
X+MP energies, which is broadened 
by the bare dephasing  of $\bar{P}$. 
Here we neglect such incoherent contributions 
to the FWM spectrum.
Similarly we approximate 
the 
photoexcited carrier density 
that determines the Pauli blocking contribution 
(first term on the rhs of Eq. (\ref{eom})) 
by the coherent exciton density,
$P^L P^{L*}$, and neglect the 
incoherent density contribution
determined by $|\bar{\psi}_1\rangle$.

The correlation function 
${\cal M}_{ii'}$ describes
the time
evolution of the long--lived MP intermediate states.
For $\sigma_+$ photoexcitation, the exciton operators 
create  spin--$\downarrow$ electrons,
and thus 
the operators $\hat{X}_i \hat{X}^\dag_j $ 
do not excite the spin--$\uparrow$ 2DEG. 
We therefore have that 
\begin{equation} 
\label{XonX}
\langle 2DEG^* |\hat{X}_i | X_j \rangle 
\sim 0, 
\end{equation} 
for any excited 2DEG state. Eq. (\ref{XonX}) 
is exact for $\nu=1$. 
Using Eq.(\ref{Y}) 
and 
Appendix \ref{MP-sym} we then derive the 
symmetry relations 
$\Omega^M_{ii'}= \Omega_M$,
${\cal M}_{10} = {\cal M}_{01}$, 
\begin{eqnarray} 
\nonumber 
( 1 - \nu_i) (1 - \nu_{j}) \,
 \langle M_{ii} | \hat{X}_j \,
| Y_{j} \rangle  = \\
- ( 1 - \nu_i) \sqrt{( 1 - \nu_j)  (1 - \nu_{j'})}
\langle M_{ii} | \hat{X}_j |Y_{j'} \rangle \nonumber \\
= - ( 1 - \nu_i) \sqrt{( 1 - \nu_j)  (1 - \nu_{j'})}
\langle M_{ii} | \hat{X}_{j'} |Y_{j} \rangle 
= W_M, \ \   \label{WM} 
\\ 
\label{MP-corr} 
 (1 - \nu_i) 
 {\cal M}_{ii} = -\sqrt{(1 - \nu_0) (1 - \nu_1) } {\cal M}_{01} 
=W_M  {\cal M}, \  \  
\end{eqnarray} 
for any  $i$  and  $j \ne j'$.
Neglecting the incoherent X+MP contribution 
to the rhs of Eq. (\ref{Ceom}), 
determined by $| \bar{\psi}_1 \rangle$,  
as compared to the first term, 
determined by the exciton polarizations,
we obtain the equation of motion 
\begin{eqnarray}
i \partial_t {\cal M} = (\Omega^{M} - i \gamma_M) 
{\cal M} \nonumber  \\
- P_{1}^L P_0^{L*} 
+ P_{0}^L P_0^{L*}
+ P_{1}^L P_1^{L*}
- P_{0}^L P_1^{L*},\ \ \ \label{M-symm}
\end{eqnarray}
where the weak MP damping, which  
to first approximation  can be  
described 
by the dephasing rate 
$\gamma_M$, \cite{marmorkos,kallin-84}
enhances the non--Markovian dephasing effects.

We note here that, in the absence of disorder, 
only the zero momentum MP state contributes 
to the nonlinear optical signal. 
As already seen in the inelastic Raman scattering spectra, 
the disorder
leads to the photoexcitation 
of a state $|M\rangle$ 
with strong contribution 
from the finite momentum 
MP's close to the 
magneto--roton energy. 
\cite{raman,marmorkos,kallin-84}
The energy $\Omega_M$ 
is 
the  average  energy of  the
coupled MP states, and  
exceeds the  cyclotron energy 
$\Omega^c_c$ that gives the zero momentum MP energy. 

Using the above results, and 
 redefining for simplicity 
$W_M \rightarrow 
W_M (1 - \nu_0) (1 - \nu_1),
$
 we obtain from Eqs. (\ref{eom}) and (\ref{Pbar}) 
the 
following closed system of equations 
for the nonlinear polarizations:
\begin{eqnarray}
i \partial_t P_0   = 
( \Omega_0 - i \Gamma_0) 
P_0
-  V_{01} ( 1- \nu_1) P_{1}^{L} \ \ \ 
\nonumber  \\
+ 
2 \mu {\cal E}(t) P_{0}^{L*} P_{0}^L
+ 
 2  V_{01} (1 - \nu_1) P_1^L P_0^L \left(  P_0^{L*} 
-  P_1^{L*}\right) \nonumber \\ 
 + W_M ( 1 - \nu_1) 
{\cal M}^* \left(P_0^{L} - P_1^L \right)
- W ( 1 - \nu_1) \bar{P}, \label{P0} 
\\
i \partial_t P_1  =
( \Omega_1 - i \Gamma_1) 
P_1 
-  V_{10} ( 1- \nu_0) P_{0}^{L} \ \ \  \nonumber \\
+ 
2 \mu {\cal E}(t) P_{1}^{L*} P_{1}^L
- 2  V_{10} (1 - \nu_0) P_1^L P_0^L \left(  P_0^{L*} 
-  P_1^{L*}\right)  
\label{P1} 
\\ - W_M ( 1 - \nu_0) 
{\cal M}^* \left(P_0^{L} - P_1^L \right)
+ W ( 1 - \nu_0) \bar{P}, \nonumber \\ 
 i \partial_t \bar{P}  =  ( \bar{\Omega} - i \gamma) 
\bar{P} + P_1 - P_0.\ \ \ 
\label{barP1}
\end{eqnarray}
The second lines in Eqs. (\ref{P0}) and (\ref{P1}) 
describe the PSF effects 
and HF XX interactions similar to the undoped 
system \cite{staff90}, 
while the third lines describe the 
correlation effects due to the time propagation of the 
intermediate Y and MP states. 
The latter correlations lead to a time--dependent  coupling 
of the LL0 and LL1 levels, as well as to non--Markovian  
dephasing.

The set of four equations
Eqs. (\ref{P0}), (\ref{P1}), 
(\ref{barP1}), and (\ref{M-symm}),  
together
with the linear polarization
equations of motion 
Eqs. (\ref{linpol-symm-0}), (\ref{linpol-symm-1}), 
and (\ref{barpol-symm}),
 constitute our  model.
To obtain the FWM spectrum, 
we assume a laser
excitation of the form $\mathcal{E}(t) =
e^{i\vec{k}_2\cdot\vec{r}}\mathcal{E}_p(t) +
e^{i\vec{k}_1\cdot\vec{r}}\mathcal{E}_p(t + \Delta t)$, where
$\mathcal{E}_p(t)$ is the Gaussian envelope of the pulses emitted
by the laser. We then solve the above equations as a function of
time $t$ and time delay $\Delta t$ between the two pulses, 
keeping only the terms leading
to a nonlinear signal in the $2\vec{k}_2-\vec{k}_1$ direction, and
perform a Fourier transform of the nonlinear polarization to get
\begin{equation} 
P(\omega,\Delta t)
= (1 - \nu_0) P_0(\omega,\Delta t)  + 
(1 - \nu_1) P_1(\omega,\Delta t).
\end{equation} 
The FWM signal measured in the experiments
is  proportional to $|P(\Delta t, \omega)|^2$
and is calculated in the next section.

\begin{figure}[t]
\begin{center}
\includegraphics*[width = 8.5cm,height=7cm]{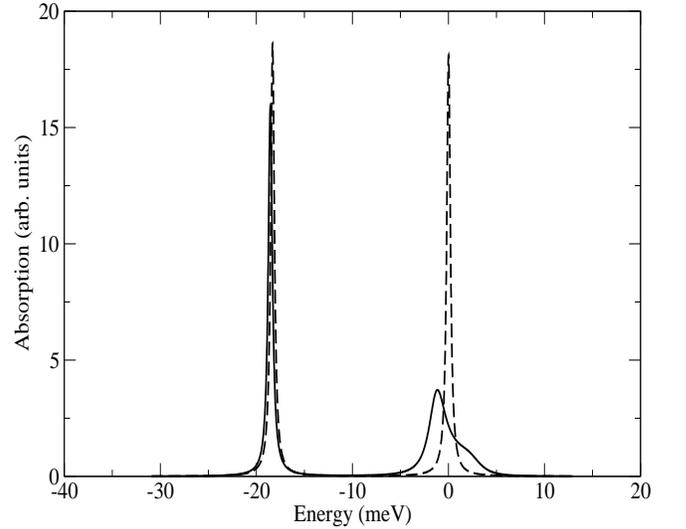}
\caption{
 Linear absorption 
spectrum (full line). 
The dashed line shows the 
spectrum 
for $\bar{P}^L$=0, in which case
the dephasing 
is determined  by the electron--phonon scattering. 
The interaction energies
$\sqrt{W}$=2.2meV and 
$V_{01}$=0.5meV, the dephasing rates
$\gamma \sim $3meV, 
and  the X and Y energies were chosen 
to reproduce 
the  linear absorption LL peak ratio, energy spacing, 
and lineshape 
observed in the experiment
of Ref. \onlinecite{from-02}
for $B$=8T and $\sigma^+$ circular polarization. In this case
$\nu_0$=0.075 and $\nu_1$=0.\cite{from-02}}
\label{lin-ab}
\end{center}
\end{figure}

\section{Numerical results} 
\label{exper} 
In this Section we present the 
results of our numerical calculations,
which are based on the model of 
section \ref{GPA}.
We start with  the linear absorption spectrum, 
$\alpha(\omega)\propto Im[P^L(\omega)/\mathcal{E}(\omega)]$.
By fitting to the
linear absorption measurements of Ref. \onlinecite{from-02}, 
we can fix the interaction 
parameters $V_{01}$ and $W$,
the  energies 
$\Omega_i$ and $\bar{\Omega}$, 
 and the dephasing
rates $\Gamma_i$ and $\gamma$, to within $\pm 50\%$.
Varying the parameters within this fitting range 
yields no significant change in the time and frequency dependence 
of the FWM or linear absorption spectrum.
Fig. \ref{lin-ab} (full line) 
displays the two X  peaks obtained this way. Their broadening is 
determined 
by (a)
the X--phonon scattering, 
described by dephasing rates 
$\Gamma_0 \sim \Gamma_1$
similar to the undoped system,
and (b) 
the  X--2DEG scattering,
described by the 
correlation function 
$\bar{P}^L$. 
The important role 
of the 
X$\rightarrow$X+MP scattering 
is clear by comparing to 
the  
dashed line curve of  Fig. \ref{lin-ab}, 
obtained with $\bar{P}^L=0$.
Although the X--2DEG scattering 
governs  the 
lineshape of the LL1  peak, 
it plays a very small role at the 
LL0 frequency.  
To interpret this behavior, 
we note that
the main 
contribution to $\bar{P}^L$ comes from 
\{1-MP + 1-LL0-e + 1-LL1-h\}
four--particle excitations
(see Eq. (\ref{Y-symm}) 
and discussion in section \ref{correl}).  
Even though $\bar{P}^L$  couples equally 
to both X amplitudes $P_0^L$ and $P_1^L$, 
it dominates the 
dephasing of $P_1^L$
since the above  four--particle 
excitations have energy comparable 
to that of $X_1$. 
In contrast, $X_0$ has significantly smaller energy, 
and thus 
the broadening of the LL0 peak is mainly determined 
by the X--phonon interactions.
Note that the asymmetric lineshape of the LL1 
resonance is due to the X+MP states and cannot be 
obtained within 
the dephasing time approximation.

\begin{widetext} 

\begin{figure}[h]
\begin{center}
\caption{
Time delay and frequency dependence of the FWM spectrum
for  excitation frequency 
(a) at the LL1  peak, 
(b) shifted by 4meV, 
(c)  shifted by 7meV,
and (d)   shifted by 9meV.
The optical pulse and linear absorption 
spectra are  displayed in the back panel.
The pulse 
duration was 150fs, 
$\Omega_M$=17.5meV, 
$\sqrt{W_M}$=2meV, and $\gamma_M$=0.2meV.}
\label{Ex} 
\end{center}  
\end{figure} 

\end{widetext}

We now study the signatures  of the X--2DEG correlations in the 
time and frequency dependence of the transient 
FWM spectrum.
As we discuss below, the correlation effects can be controlled 
experimentally by varying 
the central frequency of the optical 
pulse from LL1 toward LL0. 
This allows us to control the 
$X$ amplitudes 
$P_0$ and $P_1$,
 whose coherent superposition 
and interactions determine the 
FWM spectrum.
Figs. \ref{Ex} 
shows the effects
of such tuning. 

\begin{figure}[t]
\begin{center}
\includegraphics*[width = 8.5cm, height=7cm]{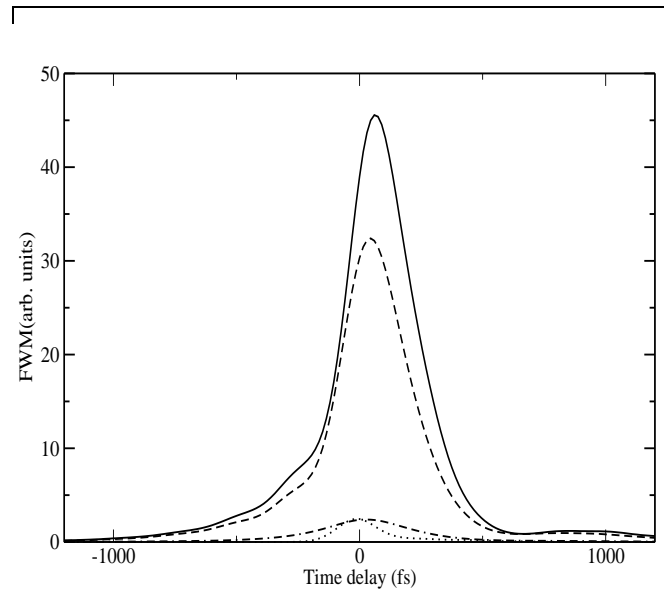}
\caption{ 
MP correlation (dashed  line),
PSF (dotted line), and XX interaction (dashed--dotted line) 
contributions to the full FWM 
signal (full line),  
calculated 
at the LL0
peak frequency for photoexcitation as in 
in Fig. \ref{Ex}(a).
 }  
\label{break0}
\end{center}
\end{figure}

\begin{figure}[t] 
\begin{center}
\includegraphics*[width = 8.5cm, height=7cm]{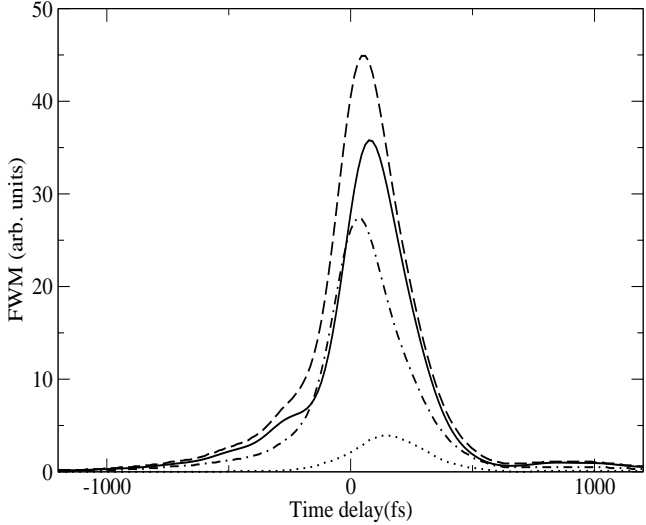}
\caption{ 
FWM signal at the 
LL0 peak frequency 
for different values of the 
average MP energy: 
$\Omega_M$=14mev (dotted line), 
17meV (full line), 18meV (dashed line), and 20meV 
(dashed--dotted line).
Photoexcitation conditions 
as in Fig. \ref{Ex}(a). 
}
\label{reson}
\end{center}
\end{figure}

Fig. \ref{Ex}(a)  
shows the FWM spectra  
when the optical pulse is centered at the 
LL1  peak, and the LL0 
peak is barely excited by the tail of the pulse.
For such excitation conditions, 
we have that $P_0^L \ll P_1^L$,
and the photoexcited density of LL1 carriers far exceeds 
that of LL0 carriers. 
As a result, 
the PSF and XX interaction contributions at the LL0 
energy are suppressed
as compared to LL1.
Despite this however, 
the LL0 and LL1 FWM peaks in Fig. \ref{Ex}(a) 
have comparable heights.  
To elucidate the physical origin of the nonresonant LL0 FWM  signal,
we show in Fig. \ref{break0}
the  contributions 
of  PSF, XX interactions, and MP correlations 
as a function of time delay.
\begin{figure}[h]
\begin{center}
\includegraphics*[width = 8.5cm, height=7cm]{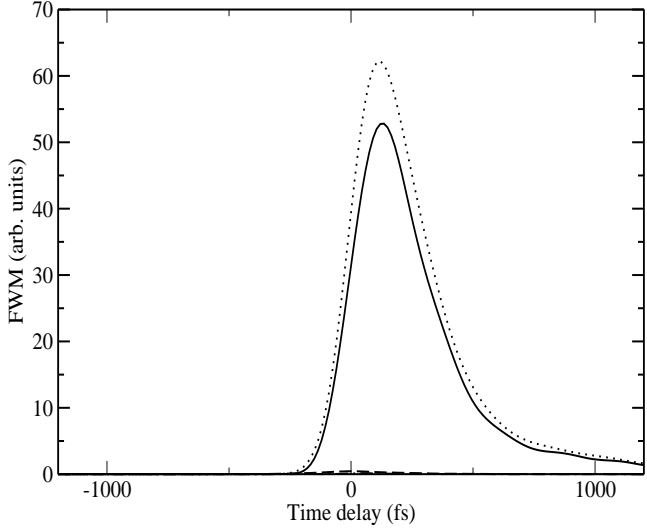}
\caption{ 
PSF (dotted line)
contribution to the  FWM 
signal (full line)
at the LL1 peak frequency
for photoexcitation as in Fig. \ref{Ex}(a).
The 
 XX interaction (dashed--dotted line) 
and MP correlation (dashed  line)
contributions 
are negligible here.
 }  
\label{break1}
\end{center}
\end{figure}
As clearly seen in the above figure, 
for optical excitation at the LL1 peak, 
the LL0 FWM signal is dominated by the MP 
correlation contribution.
At the same time, 
the LL1 signal is dominated by the PSF contribution.
This is shown in 
Fig. \ref{break1}. 
The origin  of the  strong 
MP correlation contribution to the LL0 FWM signal
can be seen by comparing the latter
for different values of 
the MP energy $\Omega_M$,
while keeping the rest of the parameters 
constant.  
As demonstrated by Fig. \ref{reson},  
the LL0 signal, 
dominated by  the MP--mediated LL coupling  
due to the process of Fig. \ref{MPC},
is {\em resonantly enhanced} as $\Omega_M$ 
approaches the 
$X_0 \rightarrow X_1$ 
excitation energy
($\sim$ 18meV here).

One should note here that 
the XX interactions also couple the two LL's.
 However, the corresponding LL0 signal 
is weaker due to  the absence of a resonance, 
similar to the MP correlation signal 
for nonresonant $\Omega_M$, and  
cannot fully account for 
the strong LL0 signal observed
 in the 
 experiment of Ref. \onlinecite{from-02}. 
To see this, note that, 
in the undoped system,
where only the XX interactions contribute, 
the FWM signal at the LL0 energy is negligible 
\cite{from-02}.
More importantly,
 in the 
 experiment of Ref. \onlinecite{from-02}, 
the LL0 peak in the doped system 
was  suppressed 
as compared to the LL1 peak 
as the density of  photoexcited carriers approached that 
of the  2DEG.
In this case the 
MP correlations of the cold 2DEG diminish, 
and the doped and undoped QW FWM signals 
start to look similar.\cite{from-02}

\begin{figure}[t] 
\begin{center}
\includegraphics*[width = 8.5cm, height=7cm]{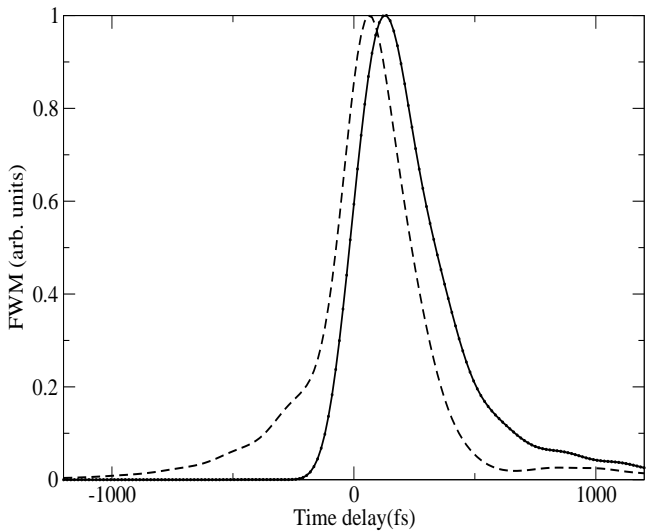}
\caption{ Temporal profile of the 
FWM spectrum at the LL0 peak frequency 
(full line) and the LL1 
peak frequency (dashed line). 
The two signals have been normalized 
for clarity.
}
\label{time}
\end{center} 
\end{figure}

We now turn to the temporal profile of the 
FWM signal.
Fig. \ref{time}, 
which plots 
the normalized time--dependent 
LL0 and LL1 signals,  
demonstrates the difference in the dynamics between 
the PSF and  MP correlation 
effects that dominate the 
LL1 and LL0 signal respectively.
As already known from undoped semiconductors, 
the Pauli blocking effects 
cannot lead to a FWM signal 
for negative time delays.
The XX  interactions lead to 
such a FWM signal,  
with  rise time   
determined by the dephasing of the 
interacting XX state $| B \rangle$.\cite{chemla-99}
Within the time dependent HF approximation, \cite{haugkoch}
the latter rise time 
is $\sim (2 \Gamma_0 + 2 \Gamma_1)^{-1}$
in the case of interest here.\cite{staff90}
For $\Gamma_0 \sim \Gamma_1$, this rise 
time 
is  about one half of the 
XX FWM decay time, $\sim (2 \Gamma_1)^{-1}$,
or the PSF FWM  decay time, 
$\sim (2 \Gamma_0)^{-1}$
at the LL0 energy.
Fig. \ref{time} however 
shows an almost symmetric
temporal profile
of the LL0 FWM signal, unlike for the LL1 
signal. 
The latter is dominated by the PSF 
contribution, and is thus suppressed 
for negative time delays, while 
the LL0 signal is dominated by the MP correlations,
and is enhanced for negative time delays, 
similar to the 
experiment of Ref. \onlinecite{from-02}. 
The  origin of this time dependence can be 
seen from the equation of motion 
of the MP correlation FWM source term in Eq. (\ref{P0}). 
After retaining only  resonant terms, one can 
see that the  rise 
of this signal 
is governed 
by the time dependence of the product $P_1^L P_0^L$,
while the decay is determined 
by the time dependence 
of $P_1^L$, ${\cal M}$, 
and by  quantum interference effects.
As discussed above, due to the X--2DEG scattering, 
$P_1^L$ dephases  much more strongly  
than $P_0^L$.
Thus
the time dependence of 
$P_1^L P_0^L$ is similar to 
that of $P_1^L$, which 
results in an almost symmetric
FWM temporal profile
at the LL0 frequency. Furthermore, 
the quantum interference and beating 
effects enhance the decay of this signal  
for positive time delays.

The relative magntitude of the MP correlation  
versus the  PSF/XX mean field 
FWM signal can be 
controlled 
experimentally by changing the  central frequency of the optical pulse.
Fig. \ref{Ex}(b)  
shows the time--dependent  FWM spectrum 
for excitation conditions 
such that $|P_0^L| \sim |P_1^L|$. 
The LL0 signal now dominates, 
and retains a temporal profile 
similar to Fig. \ref{Ex}(a). 
Note that, due to 
the increased pulse overlap with LL0,
the PSF source terms 
in Eqs. (\ref{P0}) and 
(\ref{P1})   now have comparable magnitude,
while  the XX interaction and MP correlation 
 source terms
are  also enhanced. 
However,  the 
strong dephasing 
of  $P_1$ discussed above 
suppresses the LL1 
FWM 
signal. This 
effect is magnified  in the nonlinear spectra 
as compared to the linear absorption. 
Importantly, due to the resonant enhancement 
of Fig. \ref{reson}, 
the magnitude of the 
MP correlation FWM contribution is 
enhanced more strongly 
by the increased pulse--LL0 overlap
as compared to the mean field FWM signal.
Finally, weak oscillations
as a function of time delay,
with a period 
equal to the spacing of the two LL peaks,
start to appear.

 \begin{figure}[h] 
\begin{center}
\includegraphics*[width = 8.5cm, height=7cm]{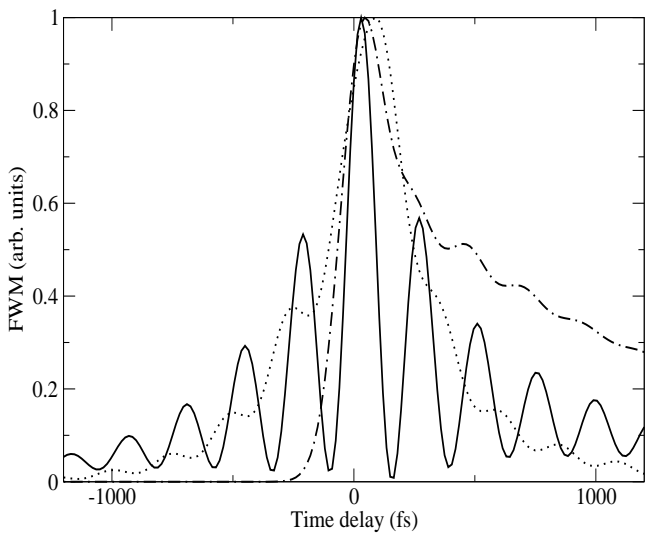}
\caption{ 
PSF (dashed--dotted line) 
and MP correlation (full line)
contributions to the  FWM 
signal 
at the LL0 frequency,
 normalized to unity for clarity, 
for photoexcitation as in Fig. \ref{Ex}(d).
We also plot the MP correlation 
signal (dotted line)  
without the LL0 coherent density 
source term of ${\cal M}(t)$ in Eq. (\ref{M-symm}).
Note the almost complete absence of oscillations 
in the latter and in the PSF contribution. 
 }  
\label{QK}
\end{center} 
\end{figure}

As can be seen  in Figs. \ref{Ex}(c) and \ref{Ex}(d),
the   oscillations of the LL0 FWM signal, 
as function of time 
{\em delay}, 
become more pronounced 
as the optical excitation frequency 
is shifted from LL1 toward LL0.
It is important to
note that the LL1 signal is almost completely 
suppressed, 
especially for the excitation 
frequency of Fig. \ref{Ex}(d),
and therefore 
there are no significant  oscillations 
in 
the {\em real time} $t$, i.e. the time related to the frequency 
$\omega$ via Fourier transform.
Thus the oscillations observed in 
Fig. \ref{Ex}(d), as well as 
in the experimental data of Ref. \onlinecite{from-02}, 
have a strong quantum kinetic contribution.\cite{weg-00,haugbook,vu-00} 
The physical origin 
of such an effect can be seen 
by plotting 
in Fig. \ref{QK} 
the PSF and MP correlation contributions to the 
LL0 signal as a function of time delay
for photoexcitation as in 
Fig. \ref{Ex}(d).
PSF leads to negligible oscillations, 
while the MP correlation 
leads to  strong oscillations.
To see the origin of the latter, 
we also plot in 
Fig. \ref{QK} the MP correlation signal 
obtained after neglecting the LL0 coherent density 
source term $P_0^L P_0^{L*}$ 
in the equation of motion 
Eq. (\ref{M-symm}) of ${\cal M}(t)$; the oscillations diminish in the latter
case. To interpret all this, 
we note that, for the excitation conditions of Fig. \ref{Ex}(d), 
we have that 
$P_0^L  \gg P_1^L$, 
and the density of LL0 carriers far exceeds that of 
LL1 carriers. 
In fact, here  
the PSF contribution 
exceeds the MP correlation contribution.
Most importantly, 
the LL0 coherent density source term 
of ${\cal M}$ 
is now  larger than the source term 
$P_1^{L} P_0^{L*}$ that gives  the 
resonant MP contribution.
Even though $P_0^{L} P_0^{L*}$ gives a nonresonant
contribution to 
${\cal M}(t)$, 
 as $P_0^L$ exceeds 
$P_1^L$
this contribution  
becomes comparable 
in  magnitude to the resonant contribution
due to 
$P_1^{L} P_0^{L*}$.
The beating between the above two resonant and nonresonant 
processes  gives rise to the  
strong oscillations.
By shifting the excitation frequency further toward LL0,  
eventually the PSF contribution dominates,
and the FWM dephasing is determined by the 
electron--phonon and  
intra--LL dephasing 
processes. \cite{from-99}

\section{Conclusions} 
\label{concl} 
In summary, we
presented a 
theory that provides a unified description
of the ultrafast 
nonlinear optical response 
of a large class 
of semiconductor 
systems with a  strongly correlated 
many--electron ground state. 
Our main result, Eq. (\ref{eom}), 
gives  the equation of motion for the third--order 
nonlinear polarization measured in 
transient wave mixing and pump--probe 
experiments, and allows us to study the 
role of the correlations and the interplay 
between coherent and incoherent effects.
Our expansion in terms of the optical 
field is valid for sufficiently  short pulses 
and/or weak excitation conditions, 
where the correlations are most pronounced. 
Our theoretical framework  allows us to describe the role 
of the long--lived collective excitations 
of a strongly correlated cold electron gas, which is  
 present prior to the optical excitation.

Our theory was applied 
to the case of the 2DEG in a strong magnetic field. 
Our numerical solution 
for photoexcitation 
close to the LL1 
energy 
with $\sigma^+-\sigma^+$ circularly polarized light 
suggests new experimental signatures 
of  collective and correlation
effects.
In this  case the relevant 2DEG 
collective excitations  
are the long--lived inter--LL magnetoplasmons, 
which  dress the 
photoexcited magnetoexcitons and lead to polaronic--like 
effects
and  strong non--Markovian dephasing. 
We showed that such effects dominate 
the time delay and frequency 
dependence  of the transient FWM
spectrum. 
FWM spectroscopy 
using femtosecond optical pulses provides both the 
time and the frequency resolution 
necessary to access this new regime 
of 2DEG physics.
Our theory allowed us to study in a systematic way 
the experimental signatures 
of the 2DEG 
quantum 
dynamics.
We  predicted,  in particular, 
a resonant enhancement of the 
lowest LL FWM signal, 
a strong
dephasing of the next 
LL magnetoexciton, 
a  symmetric FWM  temporal profile, 
and strong oscillations as function of time delay
with a strong quantum kinetic contribution.
Such predicitions agree with recent 
experimental data.\cite{from-02} 

The above correlation--induced  dynamics 
can be controlled  
by tuning the central frequency of the optical 
excitation between the two lowest LLs,
which changes the coherent admixture of the two 
MP--dressed magnetoexcitons, 
or via coherent control experiments using phase locked optical pulses.
\cite{weg-00} 
Such  experiments, 
as well as  $\sigma^-$ circularly polarized 
optical pulses,  provide new ways to 
access the very early dynamics 
of the strongly correlated  2DEG, during time scales 
shorter than the duration of the interactions. 
Such temporal and spectral resolution 
opens up new ways to 
 observe  
fractional QHE non--instantaneous correlations,  
as well as  magnon, exciton--magnetoroton,  
charged exciton, and skyrmion  effects.

\begin{acknowledgements}

We thank T.\ V.\ Shahbazyan and C. Sch\"uller for valuable
discussions, and the referee of the paper  
for helpful  suggestions. 
 This work was supported by the U.S. DoE, under
contract No. DE-AC03-76SF00098 (Berkeley), and by U.S. DoE, Grant
No. DE-FG02-01ER45916 and DARPA/SPINS (I.\ E.\ P.).

\end{acknowledgements}

\appendix

\section{}
\label{symm}
In this Appendix we derive some useful 
expressions for the operators $\hat{Y}_i$, Eq. (\ref{Yop}),
in the case of the ideal  2D  system 
displaying  electron--hole symmetry. 
To describe the magnetic field effects, 
we choose to work in the Landau gauge ${\bf A}=(0, Bx , 0)$. 
The eigenstates of the kinetic energy operator 
are then characterized by the y--component 
of the momentum, $k$, and the LL index, $n$. 
The electron, $\psi_{\alpha}$, 
and hole, $\bar{\psi}_{\alpha}$, 
eigenstates 
in this gauge are given by 
\cite{macd-85-1,shah00}
\begin{equation} 
\label{wave}
\psi_{\alpha}({\bf r})
 = \frac{e^{i k y}}{\sqrt{L}}
 \Psi_n(x-x_k) \ , \ 
\bar{\psi}_{\alpha}({\bf r})
=\psi^*_{-\alpha}({\bf r}),  
\end{equation} 
where $\alpha=(k,n,\sigma)$,
$-\alpha=(-k,n,\sigma)$,
and the 
spin--$\sigma$  wavefunction 
is kept 
implicit. In the above equation, 
$\Psi_n$ is the eigenfunction of the 1D
harmonic oscillator with frequency equal to the cyclotron frequency, 
$x_k=k l^2$ 
is the x coordinate of the cyclotron orbit center, 
$l=(\hbar c/e B)^{1/2}$ is the magnetic length (Larmor radius), 
and $L$ is the system size.\cite{macd-85-1,shah00} 

The operator $\hat{Y}_i$
is determined by 
the commutator 
$[X_i,H_{int}]$, where the  Hamiltonian 
$H_{int}=V_{ee} + V_{hh} + V_{eh}$
describes the Coulomb interactions:
\begin{widetext} 
\begin{eqnarray}
\label{H-first} H_{int}=\frac{1}{2}\int d{\bf r} d{\bf r}'
\ v({\bf r}-{\bf r}') \
\Bigl[\psi^{\dag}({\bf r})\psi({\bf r}) -\bar{\psi}^{\dag}({\bf
r})\bar{\psi}({\bf r})\Bigr]  
\Bigl[\psi^{\dag}({\bf r}')\psi({\bf r}') -\bar{\psi}^{\dag}({\bf
r}')\bar{\psi}({\bf r}')\Bigr], 
\end{eqnarray}
\end{widetext} 
where $\psi^\dag({\bf r})$ 
is the electron creation operator, 
$\bar{\psi}^\dag({\bf r})$ is the hole creation operator, 
and $v({\bf r})$ is the Coulomb potential. 
By expanding the above creation operators
in  the Landau basis
we transform the  Hamiltonian Eq. (\ref{H-first}) into the  familiar  form
\begin{widetext} 
\begin{eqnarray}
H_{int}=\frac{1}{2}\sum_{\alpha_1\alpha_2\alpha_3\alpha_4}  
\Bigl[v_{\alpha_1\alpha_2,\alpha_3\alpha_4}^{ee}
\hat{e}_{\alpha_3}^{\dag}\hat{e}_{\alpha_1}^{\dag}\hat{e}_{\alpha_2}
\hat{e}_{\alpha_4} 
+ v_{\alpha_1\alpha_2,\alpha_3\alpha_4}^{hh}
\hat{h}_{\alpha_3}^{\dag}\hat{h}_{\alpha_1}^{\dag}
\hat{h}_{\alpha_2}\hat{h}_{\alpha_4} \nonumber \\
- v_{\alpha_1\alpha_2,\alpha_3\alpha_4}^{eh}
\hat{h}_{\alpha_3}^{\dag}\hat{e}_{\alpha_1}^{\dag}\hat{e}_{\alpha_2}
\hat{h}_{\alpha_4}
- v_{\alpha_1\alpha_2,\alpha_3\alpha_4}^{he}
\hat{e}_{\alpha_3}^{\dag}\hat{h}_{\alpha_1}^{\dag}\hat{h}_{\alpha_2}
\hat{e}_{\alpha_4} \Bigr], \label{H-second} 
\ \
\end{eqnarray}
\end{widetext} 
where, in the ideal 2D system, the Coulomb interaction matrix elements
$v_{\alpha_1\alpha_2,\alpha_3\alpha_4}^{ij}$ (with $i,j=e,h$) are
given by
\begin{eqnarray}
\label{V-alpha} v_{\alpha_1\alpha_2,\alpha_3\alpha_4}^{ij}= \int 
\frac{d{\bf q}}{(2\pi)^2}v_{q}F_{\alpha_1\alpha_2}^{i}({\bf q})
F_{\alpha_3\alpha_4}^{j}(-{\bf q}),
\end{eqnarray}
where $v_{q}=2 \pi e^2/q$ is the Coulomb potential, and  
\begin{widetext} 
\begin{eqnarray}
\label{F-e} F_{\alpha_1\alpha_2}^{e}({\bf q})= \int d{\bf r}
\psi_{\alpha_1}^{\ast}({\bf r}) e^{i{\bf q}\cdot{\bf r}}
\psi_{\alpha_2}({\bf r}) \ , \ 
F_{\alpha_1\alpha_2}^{h}({\bf q})= \int d{\bf r}
\bar{\psi}_{\alpha_1}^{\ast}({\bf r}) e^{i{\bf q}\cdot{\bf r}}
\bar{\psi}_{\alpha_2}({\bf r}).
\end{eqnarray}
\end{widetext} 
Following Ref. \onlinecite{macd-85-1}
we obtain that
\begin{equation} 
\label{Fe}
F_{\alpha_1\alpha_2}^{e}({\bf q})= 
\varphi_{n_1n_2}({\bf q})f_{k_1k_2}({\bf q}) 
\delta_{\sigma_1,\sigma_2} 
\end{equation} 
where
\begin{eqnarray}
\label{f} f_{k_1k_2}({\bf q})=
e^{iq_x(k_1+k_2)l^2/2}\delta_{k_1,k_2+q_y}
\end{eqnarray}
and, for $m \ge n$, we have that 
\begin{equation}
\label{phi-mn} \varphi_{mn}({\bf q})=
\frac{n!}{m!}\biggl[\frac{(-q_y + i q_x)l}{\sqrt{2}}\biggr]^{m-n}
L_{n}^{m-n}\biggl(\frac{q^2l^2}{2}\biggr)
e^{-q^2 l^2/4}, 
\end{equation}
 where $L_n^{m-n}$ is the generalized Laguerre polynomial. 
$ \varphi_{mn}({\bf q})$ for $m < n$ can be  obtained 
by using the  property 
\begin{equation} 
\label{phi} 
\varphi_{mn}({\bf q})=
\varphi^*_{nm}(-{\bf q}).
\end{equation} 
Using Eq. (\ref{wave})
we obtain from 
Eq. (\ref{F-e}) 
\begin{eqnarray}
\label{F-h} F_{\alpha_1\alpha_2}^{h}({\bf q})= 
F_{-\alpha_2,-\alpha_1}^e({\bf q}).
\end{eqnarray}
The following symmetry relations 
can be shown by using the above relations: 
\begin{widetext} 
\begin{eqnarray} 
\label{v1} 
v_{\alpha_1\alpha_2,\alpha_3\alpha_4}^{ij}
=v_{\alpha_3\alpha_4,\alpha_1\alpha_2}^{ji}
\ , \ 
v_{\alpha_1\alpha_2,-\alpha_4-\alpha_3}^{eh}
=v_{\alpha_1\alpha_2,\alpha_3\alpha_4}^{ee} \ , \
v_{-\alpha_4-\alpha_3,\alpha_1\alpha_2}^{hh}
=v_{\alpha_3\alpha_4,-\alpha_2-\alpha_1}^{ee}.
\end{eqnarray} 
\end{widetext}
The commutator 
$[\hat{h}_{-\alpha} \hat{e}_{\alpha}, H_{int}]$ 
can be calculated from
Eq. (\ref{H-second}). 
Using Eq. (\ref{v1}) and 
some algebra we obtain that
\begin{widetext}
\begin{eqnarray}
\label{T-comm1} [\hat{h}_{-\alpha} \hat{e}_{\alpha}, H_{int}]= 
 -\sum_{\alpha_1\alpha_2}
v_{\alpha\alpha_2,\alpha_1\alpha}^{ee} \hat{h}_{-\alpha_1}
\hat{e}_{\alpha_2} +
\sum_{\alpha_1\alpha_2\alpha'} 
\  \Bigl[
v_{\alpha_1\alpha_2,\alpha\alpha'}^{ee}
\Bigl( \hat{e}_{\alpha_1}^{\dag}\hat{e}_{\alpha_2}
-\hat{h}_{-\alpha_2}^{\dag}\hat{h}_{-\alpha_1} \Bigr)
\hat{h}_{-\alpha}
\hat{e}_{\alpha'}
- (\alpha \leftrightarrow \alpha') 
\Bigr]. 
\end{eqnarray}
\end{widetext}
After  summing over
$k$, and recalling 
the definition 
Eq. (\ref{Xi-def}) 
of the X operators
 and the definition of $N_{n\sigma}$,
the lhs of the above equation 
becomes the commutator $N_{n\sigma}^{1/2} 
[\hat{X}_{n\sigma} , H_{int}]$.
Using the properties 
\begin{equation}
\label{f_prop}
\sum_k f_{kk_2}({\bf q})
 f_{k_1 k}(-{\bf q}) = \delta_{k_1k_2}
\end{equation} 
and
\begin{equation} 
\int d{\bf q} \, v(q) \phi_{nn_2}({\bf q})
 \phi_{n_1n}(-{\bf q}) 
= \delta_{n_1,n_2} 
\int d{\bf q} \, v(q) \, | \phi_{nn_1}({\bf q})|^2,
\end{equation} 
we obtain after using Eq. (\ref{Xi-def}) and some algebra 
\begin{widetext}
\begin{eqnarray}
\label{XH} 
[\hat{X}_{n\sigma}, H_{int} ] =
- \sum_{n'} \, V_{nn'\sigma}^{0} (1 - \nu_{n'\sigma}) 
\hat{X}_{n'\sigma} 
+ \frac{1}{\sqrt{N_{n\sigma}}} \,
\sum_{\alpha_1\alpha_2} 
\Bigl( \hat{e}_{\alpha_1}^{\dag}\hat{e}_{\alpha_2}
-\hat{h}_{-\alpha_2}^{\dag}\hat{h}_{-\alpha_1} \Bigr) 
\sum_{kk'n'} 
\Bigl[v_{\alpha_1\alpha_2,knk'n'}^{ee}
\hat{h}_{-kn\sigma}
\hat{e}_{k'n'\sigma}
- 
( n \leftrightarrow n') \Bigr]. \ \ \ 
\end{eqnarray} 
\end{widetext}
where 
\begin{equation}
\label{V0} 
V^{0}_{nn'\sigma} =
\frac{1}{\sqrt{(1 - \nu_{n\sigma})(1 - \nu_{n' \sigma})}}  
\int
\frac{d{\bf q}}{(2\pi)^2} \, v_{q} \, |\phi_{nn'}(q)|^2.
\end{equation}  
We  now restrict to the first two LL's,  
which dominate the optical spectra
for the excitation 
conditions of interest. 
Recalling  Eq. (\ref{Yop})
we see that the operator 
$\hat{Y}_n$ is determined by the last term 
of Eq. (\ref{XH}).
The only nonzero contribution 
to this term comes from  $n' \ne n$, 
and therefore 
$n'$=$1$ 
if $n$=0, or  
$n'$=0 if $n$=1.
As a result, 
the rhs of Eq. (\ref{XH}) 
changes sign 
between $n$=0  and $n$=1,  
and we obtain 
Eq. (\ref{Y}).
The explicit expression for the operator $\hat{Y}_{\sigma}
= \sqrt{1 - \nu_{1\sigma}} \hat{Y}_{1 \sigma}$
can then be obtained straightforwardly by subtracting 
the X contributions 
defined in   Eq. (\ref{Yop})
from the operator 
\begin{widetext}
\begin{eqnarray}
\label{Y-symm} 
\frac{1}{\sqrt{N}} \sum_{pp'kk'mm'\sigma'} 
\Bigl( \hat{e}_{pm\sigma'}^{\dag}\hat{e}_{p'm'\sigma'}
-\hat{h}_{-p'm'\sigma'}^{\dag}\hat{h}_{-pm\sigma'} \Bigr) 
\ \Bigl(v_{pmp'm',k1k'0}^{ee}
\hat{h}_{-k1\sigma}
\hat{e}_{k'0\sigma}
- 
v_{pmp'm',k0k'1}^{ee}
\hat{h}_{-k0\sigma}
\hat{e}_{k'1\sigma}
\Bigr).
\end{eqnarray} 
\end{widetext}
The subtracted  X contributions
describe corrections to the X energies and Coulomb--induced LL 
coupling 
due to the 2DEG.
As discussed in section \ref{GPA}, for 
photoexcitation
with $\sigma_+$ circularly polarized light,
we have that $\sigma=\downarrow$. 
For filling factors close to $\nu = 1$, 
the spin--$\downarrow$  states are  empty.
We can then decompose Eq. (\ref{Y-symm}) 
into the $\sigma'=\downarrow$  term, which  describes the X--X interactions,
and the $\sigma'=\uparrow$ term, 
which mainly describes  X--MP interactions.

\section{} 
\label{XX}  

In this Appendix we evaluate
the HF XX  potentials
$\langle B_{ii}  | X_{j}  X_{j'} \rangle$
in the ideal 2D system. We consider $\sigma_+$ photoexcitation
and filling factors close to $\nu=1$
so that Eq. (\ref{XonX}) applies.
Recalling the definition Eq. (\ref{B}) 
we obtain 
from 
Eq. (\ref{Y-symm}) 
after using the property 
$[\hat{X}_i, \hat{X}_j] =0$
that 
\begin{widetext} 
\begin{eqnarray}
\langle B_{1\sigma',n\sigma} | 
= \frac{1}{\sqrt{N_{n \sigma} N_{1 \sigma'}}}
\sum_{pp'kk'n'} 
\Bigl[v_{pnp'n',k1k'0}^{ee}
 \hat{h}_{-pn \sigma} \hat{e}_{p'n'\sigma}
\hat{h}_{-k1\sigma'}
\hat{e}_{k'0\sigma'} 
- v_{pn' p'n,k1k'0}^{ee} \hat{h}_{-pn'\sigma} \hat{e}_{p'n\sigma}
\hat{h}_{-k1\sigma'}
\hat{e}_{k'0\sigma'} \ \ \  
\nonumber 
 \\ 
-v_{pnp'n',k0k'1}^{ee}
\hat{h}_{-pn\sigma} \hat{e}_{p'n'\sigma}
\hat{h}_{-k0\sigma'}
\hat{e}_{k'1\sigma'}
+ v_{pn' p'n,k0k'1}^{ee}
 \hat{h}_{-pn'\sigma} \hat{e}_{p'n\sigma}
\hat{h}_{-k0\sigma'}
\hat{e}_{k'1\sigma'}
\Bigr]. \label{B-symm} 
\end{eqnarray} 
\end{widetext}
The 
only  nonzero contribution to the above equation 
comes from $n' \ne n$.
Noting 
the LL indices, 
we see that, for the conditions considered here, 
we have that  
\begin{equation} 
\label{HF1} 
\langle B_{ii} | X_j X_j \rangle 
= 0. 
\end{equation} 
Substituting 
the definition of 
$\hat{Y}_1$, 
Eq. (\ref{Yop}),
into Eq. (\ref{B}), restricting to the first two LLs, 
and denoting $i' \ne i $, 
we obtain that 
\begin{eqnarray} 
\langle B_{ii} | X_1 X_0 \rangle 
= \langle X_i X_i | H  \hat{X}^\dag_1 | X_0 \rangle
- \langle X_i| H  \hat{X}_i |X_1 X_0 \rangle \ \ \
\nonumber
 \\
- \Omega_i  \langle X_i X_i |X_1 X_0 \rangle
+
V_{ii'}  \langle X_{1}  X_{0} |X_1 X_0 \rangle
- \langle Y_i | X_i | X_1 X_0 \rangle. \ \ 
\end{eqnarray} 
We have that 
$\langle X_i  X_i |X_1 X_0 \rangle=
\langle X_i  X_i |X_{i'} X_{i'} \rangle=
0$
due to the orthogonality of the valence hole states, while 
$\langle Y_i | \hat{X}_i | X_1 X_0 \rangle = 0$ 
due to Eq. (\ref{XonX}). 
Using the above, Eq. (\ref{Yop}) 
for 
 the commutator 
$[ H , \hat{X}^\dag_1]$,  Eq. (\ref{HonX}) 
for the states $H | X_0 \rangle$ 
and  $\langle X_i| H$, 
and 
Eq. (\ref{HF1}), 
we obtain 
after some algebra
that 
\begin{eqnarray} 
\langle B_{ii} | X_1 X_0 \rangle =
 2 V_{ii'}  \langle X_1 X_0 |X_1 X_0 \rangle \nonumber
\\
- V_{01} \langle X_i X_i |   X_0 X_0\rangle
- V_{10} \langle X_i X_i |   X_1 X_1 \rangle.
\end{eqnarray} 
Using the relations 
$\langle X_0 X_1 |X_1 X_0 \rangle=1$
and 
\begin{equation}
\langle X_i X_i | X_i X_i \rangle
=2(1 - \frac{1}{N_i}),
\end{equation}
obtained from Eq. (\ref{commut}), 
we finally obtain that 
\begin{equation}
\label{VXX}  
N_i
\, \langle B_{ii} | X_1 X_0 \rangle =
2 V_{ii'} \ , \ i' \ne i.
\end{equation} 
The above relation recovers the results of Ref. \onlinecite{staff90}.

\section{} 

\label{MP-sym}

In this Appendix we derive some useful relations for the 
overlap $\langle \rm{2DEG^*}| M_{ii'} \rangle$, 
where $| 2DEG^* \rangle$ is any 2DEG excited state, 
for filling factors close to $\nu=1$ and 
for $\sigma_+$ polarized light.
Using Eqs. (\ref{Mstate}) and  (\ref{Y})
we  obtain that 
\begin{eqnarray} 
N^{1/2}_0
| M_{0i} \rangle 
= N^{1/2}_0
 \hat{Y}_0 \hat{X}^\dag_i | 0\rangle  \nonumber \\
= -N^{1/2}_1   \hat{Y}_1 \hat{X}^\dag_i | 0\rangle
= - N^{1/2}_1   | M_{1i} \rangle.
\label{M-symm-1} 
\end{eqnarray} 
From Eq. (\ref{Mstate}) 
we obtain after using Eq. (\ref{Yop}) 
that 
\begin{eqnarray} 
|M_{ii'} \rangle = \hat{X}_i | Y_{i'} \rangle 
- (H + \Omega_i - \Omega_{i'}) \hat{X}_i | X_{i'} \rangle 
\nonumber 
\\
+ \sum_{j \ne i} V_{ij} \hat{X}_j | X_{i'} \rangle 
-  \sum_{j \ne i} V_{ji'} \hat{X}_i | X_{j} \rangle.
\end{eqnarray} 
The state $ \hat{X}_i | Y_{i'} \rangle$
describes a 2DEG excitation, 
created via the process shown in the first 
three panels of Fig. (\ref{MPC}). 
Using Eq. (\ref{XonX}) 
and the property   $ \langle 2DEG^* | H | 0 \rangle=0$
we obtain 
that 
\begin{equation} 
\label{XYYX}
\langle 2DEG^* | M_{ii'} \rangle 
=\langle 2DEG^* |\hat{ X}_i | Y_{i'} \rangle.
\end{equation} 
Using Eq. (\ref{Y})
we then obtain 
that 
\begin{eqnarray} 
N_0^{1/2} \langle \rm{2DEG^*}| M_{i0} \rangle 
= N_0^{1/2} \langle \rm{2DEG^*} |\hat{X}_i |Y_0 \rangle \nonumber 
\\ 
= - N_1^{1/2} \langle \rm{2DEG^*} |\hat{X}_i |Y_1 \rangle 
=-N_1^{1/2}\langle \rm{2DEG^*} | M_{i1} \rangle.
\label{M-symm-2} 
\end{eqnarray} 
%



\end{document}